
%
%
%
%
%
%
%
%
\input phyzzx
\NPrefs
\def\dir{}     
\def\figcond{1}     
\def\figsty{0}      
%
\newcount\figsubcountss \figsubcountss=0
\ifnum\figcond>0 \input epsf 
  \ifnum\figsty>0 \openout10=\jobname.fis \fi
\fi
\def\figinsert#1#2#3{
\ifnum\figcond>0 
  \ifnum\figsty>0 
    \global\advance\figsubcountss by 1
    \ifnum\figsubcountss>12
      \ifnum\figsubcountss<14
        \immediate\write10{\noexpand\  \noexpand\vskip10pt}
      \else
        \ifodd\figsubcountss
          \immediate\write10{\noexpand\  \noexpand\vskip10pt%
                             \noexpand\vfill}
        \else
          \immediate\write10{\noexpand\  \noexpand\vskip10pt}
        \fi
      \fi
    \else
      \ifodd\figsubcountss
        \immediate\write10{\noexpand\  \noexpand\vskip10pt}
      \else
        \immediate\write10{\noexpand\  \noexpand\vskip10pt%
                           \noexpand\vfill}
      \fi
    \fi
    \immediate\write10{\noexpand\fbox{Figure \noexpand#1}}
    \immediate\write10{\noexpand\vskip60pt}
    \immediate\write10{\noexpand\centerline{\noexpand\epsfbox{\dir#2}}}
    \ifnum\figsubcountss>12
      \ifnum\figsubcountss<14
        \immediate\write10{%
                  \noexpand\vskip80pt\noexpand\vfil\noexpand\eject}
      \else
        \ifodd\figsubcountss
          \immediate\write10{%
                    \noexpand\vskip80pt\noexpand\vfil\noexpand\eject}
        \else
          \immediate\write10{\noexpand\vskip 50pt \noexpand\vfill}
        \fi
      \fi
    \else
      \ifodd\figsubcountss
        \immediate\write10{\noexpand\vskip 50pt \noexpand\vfill}
      \else
        \immediate\write10{%
                  \noexpand\vskip80pt\noexpand\vfil\noexpand\eject}
      \fi
    \fi
  \else
    \midinsert
    \vskip 10pt
    \centerline{\epsfbox{\dir #2}} \vskip9pt
  \fi
\else
  \ifnum\figsty=0 \midinsert \fi
\fi
\ifnum\figsty>0 \FIGURE#1{#3} 
\vskip15pt\noindent\fbox{Fig. #1}\vskip15pt\noindent
\else \captionbox#1{#3} \endinsert
\fi
}
\def\figepsfout{\Closeout10
\nopagenumbers
\input \jobname.fis}
\newbox\tempboxa
\newdimen\captionboxsubcount \captionboxsubcount=30pt
\newdimen\captionboxsub
\def\captionbox#1#2{\FIGNUM #1 %
\ifnum\figcond=0 \fbox{Figure #1} \fi
\captionboxsub=\hsize \advance\captionboxsub by -\captionboxsubcount
\setbox\tempboxa\hbox{Fig. #1. #2}
\ifdim \wd\tempboxa >\captionboxsub 
{\narrower {\narrower \hangafter=1 
\ifnum\figurecount>9 \hangindent=50pt
\else \hangindent=40pt \fi
\noindent{Fig.~#1.} #2\par}} 
\else \hbox to\hsize{\hfil\box\tempboxa\hfil}\fi}

\def\fboxsub#1{\vbox{\hrule\hbox{%
\vrule\kern3pt\vbox{\kern3pt#1\kern3pt}\kern3pt\vrule}\hrule}}
\def\fbox#1{\setbox4=\vbox{\kern3pt\hbox{#1}\kern3pt}
\hfill\fboxsub{\box4}\hfill\ \par}
\def\ee{\eqno\eq }
\def\abs#1{{\left\vert #1 \right\vert}}
\def\Sf{S_{\rm F}}           
\def\PVGSf{{\cal S}_{\rm F}}     
\def\Sfinv{{S_{\rm F}^{-1}}}     
\def\PVGSfinv{{{\cal S}_{\rm F}^{-1}}}     
\def\kterm{{\cal K}_{\rm 2PI}}  
\def\Tr{{\rm Tr}}
\def\tr{{\rm tr}}
\def\Ln{{\rm Ln}}
\def\chihat{\hat{\chi}}
\def\slash#1{{#1\llap{\it/}}}
\def\Dslash{{D\kern-9pt\hbox{\it/}\kern3.5pt}}
\def\delslash{\partial\kern-7.5pt\hbox{\it/}\kern3pt}
\def\Aslash{{A\kern-7.5pt\hbox{\it/}\kern3.5pt}}
\def\kslash{{k\kern-7.5pt\hbox{\it/}\kern3.5pt}}
\def\del{\partial}
\def\delm{\partial_\mu}
\def\deln{\partial_\nu}
\def\Ktil{\tilde{K}}
\def\pigen{{\tau_3\over2}}
\def\AnomFac{{ N_{\rm c} \over 4 \pi^2 } \tr\left( \tau_3 Q Q \right)%
\epsilon^{\mu\nu\rho\sigma} p_\rho k_\sigma }
\def\PR#1#2#3{{ Phys. Rev. }{\bf #1} {(#3)} #2}

\def\PL#1#2#3{{ Phys. Lett. }{\bf #1} {(#3)} #2}

\def\NC#1#2#3{{ Nuovo Cim. }{\bf #1} {(#3)} #2}

\def\PROG#1#2#3{{ Prog. Theor. Phys. }{\bf #1} {(#3)} #2}

\def\JMATH#1#2#3{{J. Math. Phys.} {\bf #1} {(#3)} #2}
\def\REFSUB#1#2#3{{\bf #1} {(#3)} #2}
\REF{\refBSreva}{
For a recent review of these methods, see, for example, T.~Kugo, in 
{\it Proc. of 1991 Nagoya Spring School of Dynamical Symmetry
Breaking}, Apr. 23--27, 1991, ed. K.~Yamawaki (World Scientific Pub.
Co., Singapore, 1991).}
\REF{\MN}{
T.~Maskawa and H.~Nakajima, \PROG{52}{1326}{1974};
\REFSUB{54}{860}{1975}.}
\REF{\refKugoMitchard}{
T.~Kugo and M.G.~Mitchard, \PL{B282}{162}{1992};
\PL{B286}{355}{1992}.}
\REF{\refABJ}{
S.L.~Adler, \PR{177}{2426}{1969}:
J.S.~Bell and R.~Jackiw, \NC{60A}{46}{1969}.}
\REF{\refNJL}{
Y.~Nambu and G.~Jona-Lasinio, \PR{10}{345}{1961}.}
\REF{\refDM}{
C.~De~Dominicis and P.C.~Martin, \JMATH{5}{14}{1964};
\REFSUB{5}{31}{1964}.}
\REF{\refCJT}{
J.M.~Cornwall, R.~Jackiw and E.~Tomboulis, \PR{D10}{2428}{1974}.}
\REF{\Jackiw}{
R.~Jackiw, in {\it Lectures on Current Algebra and its Applications},
by S.B.~Treiman, R.~Jackiw and D.J.~Gross,
(Princeton Univ. Press, 1972).}
\REF{\refAdlerBardeen}{
S.L.~Adler and W.A.~Bardeen, \PR{182}{1517}{1969};
W.A.~Bardeen, \PR{184}{1848}{1969}.}
\REF{\refGSC}{
H.~Georgi, E.H.~Simmons and A.G.~Cohen, \PL{B236}{183}{1990}.}
\REF{\refPagelsStokar}{
H.~Pagels and S.~Stokar, \PR{D20}{2947}{1979}.}
\date{%
KUNS-1236 \cr
HE(TH) 93/13 \cr
hep-ph/9312343 \cr
December, 1993}

\titlepage
\title{External Gauge Invariance and Anomaly \break
in BS Vertices and Boundstates}
\author{Masako Bando, Masayasu Harada\foot{
Fellow of the Japan Society for the Promotion of Science 
for Japanese Junior Scientists.}
and Taichiro Kugo}
\address{
Department of Physics,
Kyoto University\break
Kyoto 606, Japan
}
\abstract{
A systematic method is given for obtaining consistent approximations to 
the Schwinger-Dyson(SD) and Bethe-Salpeter(BS) equations which 
maintain the external gauge invariance. 
We show that 
for any order of approximation to the SD equation there is a 
corresponding approximation to the BS equations such that 
the solutions to those equations satisfy the 
Ward-Takahashi identities of the external gauge symmetry.
This formulation also clarifies the way how we can calculate the 
Green functions of current operators in a consistent manner 
with the gauge invariance and the axial anomaly. 
We show which type of diagrams for 
the $\pi^0\rightarrow\gamma\gamma$ amplitude using 
the pion BS amplitude give result 
consistent with the low-energy theorem. 
An interesting phenomenon is observed in the ladder approximation 
that the low energy theorem is saturated by the zeroth order terms 
in the external momenta of 
the pseudoscalar BS amplitude and the vector vertex functions.
}
\endpage

\chapter{Introduction}

If we consider a strongly interacting fermion system, 
we have to deal with various boundstates and their interactions 
among themselves or to some external gauge fields. For instance, 
in QCD, the boundstates are hadrons, which are in fact the only 
observable states, and the external gauge fields are the photon 
and the $W$ and $Z$ bosons. In order to treat such 
boundstate problems, 
we are obliged to adopt some {\it non-perturbative} approximation 
scheme, and then it becomes a non-trivial issue whether and how to 
keep the external gauge invariance. 

In this paper we discuss this problem for the case using 
the Schwinger-Dyson (SD) and Bethe-Salpeter (BS) equations 
as a non-perturbative approximation method\refmark{\refBSreva}. 
We use, in a certain approximate manner, 
the SD equation for calculating fermion propagator and the 
inhomogeneous BS equations for the vertices of fermion 
with the external gauge bosons. 
Then the adopted approximation has to keep the external gauge 
invariance in order to give results consistent with, for instance, 
various low-energy theorems.  
This problem is actually not so trivial and, in fact, 
it will turn out that the approximations for the SD and BS 
equations cannot be independent of one another 
but have to satisfy a {\it mutual consistency}. This is 
because the fermion propagator and the vertex functions are connected 
to one another by the Ward-Takahashi (WT) identities resulting from 
the external gauge invariance. 
We shall present a general answer to this problem and find that 
{\it for any order of approximation to the SD equation there is a 
corresponding approximation to the BS equations which satisfies the 
external gauge invariance}. 

A partial answer to this problem has actually been known 
to Maskawa and Nakajima\refmark{\MN,\refKugoMitchard} 
for a long time: the axial-vector vertex function satisfies the 
axial-vector WT identity if it is determined by the ladder BS equation 
in which the fermion propagator determined by the ladder SD equation 
is used. They proved this result by a direct diagrammatic 
evaluation of the divergence of the axial-vector vertex in that 
approximation. Our result in this paper is to give a generalization to 
arbitrary order of approximations. The proof of the gauge invariance 
is given in a very systematic manner. 

Gauge invariance gives constraints not only on those 
fermion propagator and vertex functions 
but also the Green functions of the vector or axial-vector
current operators coupled to the external gauge fields. 
We also discuss such Green functions of current operators 
in the same framework. We consider, for instance, 
a 3-point Green function of one axial-vector current and
two vector currents, which gives a well known anomaly and 
is related the $\pi^0\rightarrow\gamma\gamma$ amplitude by
the low-energy theorem\refmark{\refABJ}.
In the ladder approximation, as we show, 
the Green function simply obtained by a triangle diagram in which 
the vertices and fermion propagators are those in that approximation, 
satisfies the vector gauge invariance and reproduces the correct 
axial anomaly. 
This simple situation is, however, no longer true 
in any approximation beyond the ladder. We shall show that 
we have to include corrections intrinsic to the Green function which 
can be attributed neither to the fermion propagator nor to 
the vertex functions. 
This is also the case, even in ladder approximation, for general 
$n$-point Green function of current operators with $n\geq4$.  
For instance, consider the 4-point Green function of vector currents 
corresponding to photon scattering 
$\gamma\gamma\rightarrow\gamma\gamma$. 
The simple box diagram consisting of the vertices and fermion 
propagators calculated in the ladder approximation does not satisfy 
gauge invariance, despite that 
the vertex function and the propagator themselves 
are mutually consistent with WT identity in the ladder approximation. 
We shall clarify in a general manner which types of diagrams should be 
included in order to obtain gauge-invariant Green functions (with 
correct anomaly in case it is present).

We discuss, in particular, the calculation of
$\pi^0\rightarrow\gamma\gamma$ decay amplitude by using pseudoscalar
BS amplitude in detail, 
since it gives a typical example in which all these problems of 
gauge invariance and anomaly become relevant. 
We show that the resultant amplitude satisfies the low-energy theorem
if the boundstate BS amplitude is the one determined by the BS
equation in the `same' order of approximation as those for the fermion
propagator and the vector vertices, and if we include a particular set
of diagrams which depend on the approximation.
We also point out an interesting phenomenon in the ladder approximation 
that the low energy theorem is saturated by the zeroth order terms in 
the external momenta of 
the vector vertices and the pseudoscalar BS amplitude.

This paper is organized as follows.
In section 2, we introduce a gauge invariant effective action for 
the fermion propagator on arbitrary background of external gauge 
fields. This provides us with a very systematic way to 
obtain mutually consistent SD and BS equations. 
It is proved that the solutions to these equations satisfy 
the WT identities in section 3. 
In section 4, we construct 3-point 
Green function of one axial-vector and two vector current operators,
and show that it not only satisfies the vector gauge invariance but 
also reproduces correct axial anomaly. 
Section 5 is devoted to consideration of the pseudoscalar boundstate
and the low-energy theorem for
$\pi^0\rightarrow\gamma\gamma$. We show which types of diagrams give 
amplitude consistent with the low-energy theorem. 
In section 6, we discuss the above mentioned phenomenon of 
zeroth order saturation of the low-energy theorem 
in the ladder approximation. An appendix is added to show 
how the discussion of the anomaly in section 4 goes 
when we adopt dimensional 
regularization instead of the Pauli-Villars-Gupta's one adopted in 
the text.

\chapter{Gauge Invariant Effective Action}

We consider an interacting fermion system like QCD in which the chiral 
symmetry is spontaneously broken dynamically.  As usual we call the 
`gauge' interaction responsible for the formation of boundstates and 
for the spontaneous chiral symmetry 
breaking ``color gauge interaction'' and the gauge boson ``gluon'',
although the present formulation also applies to more general systems
than QCD.\foot{
The `gauge' boson need not be a true gauge boson; so, for example, 
it may have non-zero masses and may be axial-vector. 
Then the Nambu-Jona-Lasinio\refmark{\refNJL} 
like models can be discussed by considering
a limit the vector and axial-vector `gauge' boson masses go to very 
large.}
In such a system we want to calculate, in an {\it approximate} 
but {\it non-perturbative} manner, the fermion propagator as well as 
the (color-singlet) vector and axial-vector vertices to which 
external gauge fields couple. A method is to use the Schwinger-Dyson 
(SD) equation for the fermion propagator and the inhomogeneous 
Bethe-Salpeter (BS) equation for the vertices. 
In this case, as announced in the Introduction, 
the approximations adopted for the SD and BS 
equations cannot be independent one another but have to satisfy 
a mutual consistency in order to meet the 
the external gauge invariance requirement. 

A systematic way to obtain those SD and BS equations satisfying the 
mutual consistency, is provided if we consider the original system 
put in a general external gauge field background. Let us denote the 
external background gauge fields as  
$A_\mu \equiv A_\mu^a\lambda ^a$ with flavor matrix 
$\lambda ^a$ normalized as
$\tr(\lambda ^a\lambda ^b)=(1/2)\delta _{ab}$, 
and assume a vector-like coupling to 
the fermion: ${\cal L}_{\rm int}=\bar\psi \gamma _\mu A^\mu \psi $. 
We assume this just 
for notational simplicity and axial-vector case can be obtained simply 
by replacement $\gamma _\mu  \rightarrow  \gamma _\mu \gamma _5$. An 
important assumption is 
that the flavor degrees of freedom to which the external gauge fields 
couple are {\it orthogonal} to the color degrees of freedom to 
which the `gluons' (internal `gauge' bosons) couple.

Effective action for fermion propagator  $\Sf$ in the presence of
external gauge field background is given by\refmark{\refDM,\refCJT}
$$
  \Gamma[\Sf,A] = i \Tr \, \Ln \Sf - \Tr
  \left( i \Dslash \Sf \right)
  + i^{-1}\kterm [ \Sf] ,
\eqn{\eqEfAc}
$$
where we note that
the external gauge field $A_\mu$ is present only at the covariant
derivative $D_\mu=\delm-iA_\mu$.
Here
$\kterm$ stands for the two particle irreducible (w.r.t. fermion-line)
diagram contributions:
for example in QCD, 
we can expand the $\kterm$ 
into power series of the gauge coupling $\alpha_s = g_s^2/4\pi$,
$$
  \kterm = \kterm^{(1)}
  + \kterm^{(2)} + \cdots
\eqno\eq
$$
and 
$\kterm^{(1)}$ and $\kterm^{(2)}$
are diagrammatically given by Fig.~1.
\figinsert\figefac{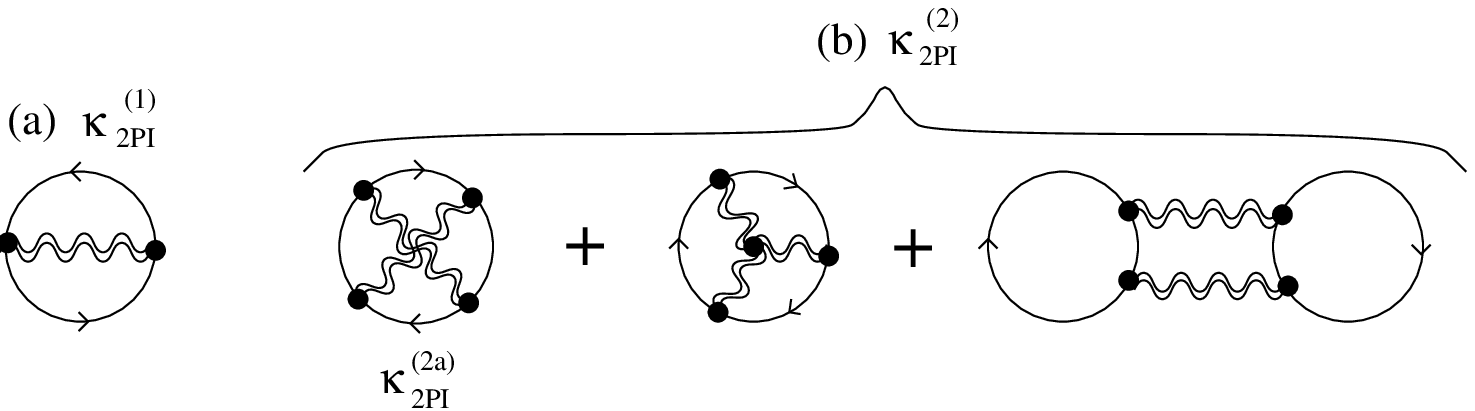}%
{Two particle irreducible (w.r.t. fermion-line) diagrams contributing 
to $\kterm^{(1)}$ and $\kterm^{(2)}$ in QCD.
The double wavy line represents the gluon propagator $D_{\mu \nu }$ 
and the solid line represents the fermion propagator $\Sf$.}
More explicitly the first term $\kterm^{(1)}$ is given by\foot{
If we use the running coupling constant, the coupling constant $g_s$
depends on the gluon momentum and hence we should understand that 
the running coupling factors $g_s^2$ in such a case 
are placed at the loop integrand.
All the discussions in this paper apply equally both to the fixed and
running coupling cases, provided that in the latter case the gluon
momentum is used as the argument of the running coupling 
function\refmark{\refKugoMitchard}.} 
$$
  \kterm^{(1)} =
  - {g_s^2 \over 2}
  \int d^4x d^4y\ \tr
  \left(
    \Sf(x,y) i \gamma_\mu T^a \Sf(y,x) i \gamma_\nu T^a
  \right)
  D^{\mu\nu}(x-y) \ ,
\ee
$$
where $T^a \ (a=1,\cdots ,N_c^2-1)$ are color matrices in the quark 
representation and $D_{\mu \nu }$ is tree level gluon propagator
given by
$$
  D^{\mu\nu}(x) = \int {d^4p\over i(2\pi )^4} e^{-ipx}
  \left(
    g^{\mu\nu} - { p^\mu p^\nu \over p^2}
  \right)
  \left(
    { 1\over p^2 } - { 1\over p^2 - \Lambda^2 }
  \right) 
  \qquad \quad 
  ( \Lambda \rightarrow \infty ) \ ,
\eqn\eqREGGLUON
$$
where we have included an ultraviolet cutoff $\Lambda $ for 
definiteness. 
For later convenience, we refer to the first diagram 
contribution to the $\kterm^{(2)}$
in Fig.~\figefac(b) as $\kterm^{(2a)}$:
$$
\eqalign{
  \kterm^{(2a)} 
& =
  - {g_s^4  \over 4}
  \int d^4x_1 d^4x_2 d^4y_1 d^4y_2 \ 
  D^{\mu\nu}(y_1-y_2) D^{\rho\sigma}(x_1-x_2) .
\cr
& \qquad
  \times
  \tr
  \left(
    \Sf(x_1,y_1) i \gamma_\mu T^a \Sf(y_1,x_2) i \gamma_\rho T^b
    \Sf(x_2,y_2) i \gamma_\nu T^a \Sf(y_2,x_1) i \gamma_\sigma T^b
  \right) .
\cr
}
\eqno\eq
$$

By our assumption that the flavor is a freedom orthogonal 
to the color, we note that the flavor matrices $\lambda ^a$ commute with
the color matrices $T^a$. Then we have the following lemma:

\noindent
{\bf Lemma:}\ {\it For any approximation for $\kterm$ by an arbitrary 
subset of diagrams contributing to $\kterm$, the 
effective action Eq.\eqEfAc\ is (external) gauge invariant}:
$$
  \Gamma[\Sf,A] = \Gamma[\Sf^U,A^U] \ ,
\eqno\eq
$$
{\it where the gauge transformation with 
$U(x)=\exp\big(i\theta ^a(x)\lambda ^a\big)$ 
is given explicitly by}
$$
\eqalign{
  A_\mu 
& \rightarrow
  A^U_\mu  = U A_\mu  U^{-1} + {1\over i} \del_\mu  U \cdot U^{-1} \ ,
\cr
  \Sf(x,y)
& \rightarrow
  \Sf^U(x,y) = U(x) \Sf(x,y) U^{-1}(y) \ .
}
\eqn\eqGAUGETR
$$

\noindent
Proof) 
It is convenient to introduce the following functional notation:
$$
\eqalign{
  \left( U \right)_{xy}
& \equiv
  U(x) \delta^4 (x-y) \ ,
\cr
  \left( \Dslash \right)_{xy}
& \equiv
  \left( \delslash _x - i \Aslash (x) \right) 
  \delta^4 (x-y) \ ,
\cr
  \left( \Sf \right)_{xy}
& \equiv
  \Sf (x,y) \  .
\cr
}
\eqno\eq
$$
In this notation we can rewrite the gauge transformation \eqGAUGETR\ 
simply in the form
$$
\eqalign{
  \Dslash
& \rightarrow 
  U \Dslash U^{-1} \ ,
\cr
  \Sf
& \rightarrow 
  U \Sf U^{-1} \ .
\cr
}
\eqno\eq
$$
Then we can see the gauge invariance of the first two terms in $\Gamma 
$ 
as follows:
$$
\eqalignno{
  \Tr \Ln \Sf 
& \rightarrow 
  \Tr \Ln \left( U \Sf U^{-1} \right)
  = \Tr \left[ U \left( \Ln \Sf \right) U^{-1} \right]
  = \Tr \Ln \Sf \ ,
& \eq
\cr
  \Tr \left( \Dslash \Sf \right) 
& \rightarrow
  \Tr 
  \left(
    U \Dslash U^{-1} U \Sf U^{-1}
  \right)
  = \Tr \left( \Dslash \Sf \right) \ .
& \eq
\cr
}
$$
The gauge invariance of $\kterm[\Sf]$ is also shown similarly: for 
instance, for the lowest order diagram we have
$$
\eqalign{
  \kterm^{(1)}
& \rightarrow
  - {g_s^2 \over 2}
  \int d^4x d^4y \ \tr
  \left(
    U(x) \Sf(x,y) U^{-1}(y) i \gamma_\mu T^a U(y) \Sf(y,x) U^{-1}(x) 
    i \gamma_\nu T^a 
  \right)
\cr
& \qquad\quad
  \times
  D^{\mu\nu}(x-y) 
  = \kterm^{(1)} \ .
\cr
}\ee
$$
The point here is that the fermion propagators appear successively 
in the trace and the vertices there contain only color and $\gamma $ 
matrices 
which commute with the flavor gauge transformation matrices $U(x)$. 
This property clearly holds for any diagrams contributing to $\kterm$ 
and so the gauge invariance follows. This finishes the proof.

Because of the lemma,
the Schwinger-Dyson (SD) equation derived from the action $\Gamma$ is
automatically (external) gauge covariant 
(or gauge invariant as a set of equations).

The SD equation is given by 
$\delta \Gamma /\delta \Sf=0$, which reads  
$$
  i \Sf^{-1} = i \Dslash - i^{-1}
  { \delta \kterm \over \delta \Sf } \ .
\eqn\eqSDeq
$$
If we take only the lowest order term in $\kterm$, $\kterm^{(1)}$, 
then this SD equation reduces to
(see Fig.~2) 
$$
  i \Sf^{-1} = i \delslash  + \Aslash  
  + i^{-1}K \ast \Sf \ ,
\eqn\eqSDeqlowest
$$
with $K\ast\Sf$ defined by
$$
  K \ast \Sf
  \equiv  g_s^2
   (i \gamma_\mu T^a) \Sf(y, x) (i \gamma_\nu T^a) D^{\mu\nu}(x-y)\ .
\eqn\eqDefKop
$$
\figinsert\figSDeq{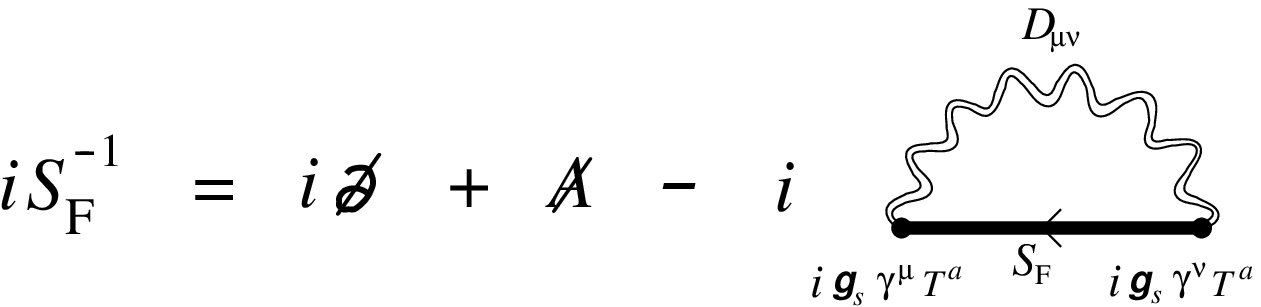}{Schwinger Dyson equation derived from
the effective action $\Gamma$ using $\kterm=\kterm^{(1)}$.}
\noindent
Eq.\eqSDeq\ (or \eqSDeqlowest) is the SD equation determining 
a solution $\Sf=\Sf[A]$ for the fermion propagator, 
on an {\it arbitrary} external 
background gauge field $A_\mu$.

The solution
$\Sf[A]$ is expanded into a power series in 
the external gauge field $A_\mu$:
$$
  \Sf[A] =
  \Sf + i A_\mu^a G_3^{a\mu}
  + { i^2 \over 2 } A_\mu^a A_\nu^b G_4^{a\mu,b\nu}
  + { i^3 \over 3 } A_\mu^a A_\nu^b A_\rho^c G_5^{a\mu,b\nu,c\rho}
  + \cdots \ ,
\eqn\eqSfExpand
$$
where $a$, $b$ and $c$ denote the flavor indices.
Here and henceforth the space-time coordinates and the 
integrations are suppressed,
i.e., $A_\mu^a G_3^{a\mu} \equiv 
\int d^4z A_\mu^a(z) G_3^{a\mu}(x,y;z)$, etc.. 
The function
$G_{n+2}^{a_1\mu_1,\cdots,a_n\mu_n} (x,y;z_1,\cdots,z_n)$
defines a fermion 2-point function with $n$ vector vertices 
inserted (see Fig.~3):
$$
\eqalign{
  G_3^{a\mu} (x,y;z)
&\equiv
  \left.
    {1 \over i}
    { \delta \Sf(x,y;A) \over \delta A_\mu^a(z) }
  \right\vert_{A=0} 
  =
  \bra{0}
    {\rm T} j^{a\mu}(z) \psi(x) \bar{\psi}(y)
  \ket{0} \ ,
\cr
  G_4^{a\mu,b\nu} (x,y;z,w)
&\equiv
  \left.
    {1 \over i^2}
    { \delta \Sf(x,y;A) \over \delta A_\mu^a(z) \delta A_\nu^b(w)}
  \right\vert_{A=0} 
  =
  \bra{0}
    {\rm T} j^{a\mu}(z) j^{b\nu}(w) \psi(x) \bar{\psi}(y)
  \ket{0} , \ 
  \ldots \, .
\cr
}
\eqn\eqGF
$$
This is because $\delta /\delta A_\mu ^a$ yields an insertion of the 
vector current 
operator $j^{a\mu}=\bar\psi \gamma ^\mu \lambda ^a\psi $ 
to which the external gauge boson $A_\mu ^a$ couples.
Hereafter in this section, we suppress the flavor indices to denote 
$G_{n+2}^{a_1\mu_1,\cdots,a_n\mu_n}$ simply as
$G_{n+2}^{\mu_1\cdots\mu_n}$,
and write only $\gamma^\mu$ in place of $\gamma^\mu\lambda^a$ as
vertex factors in the figures, accordingly.
\figinsert\figDefofThree{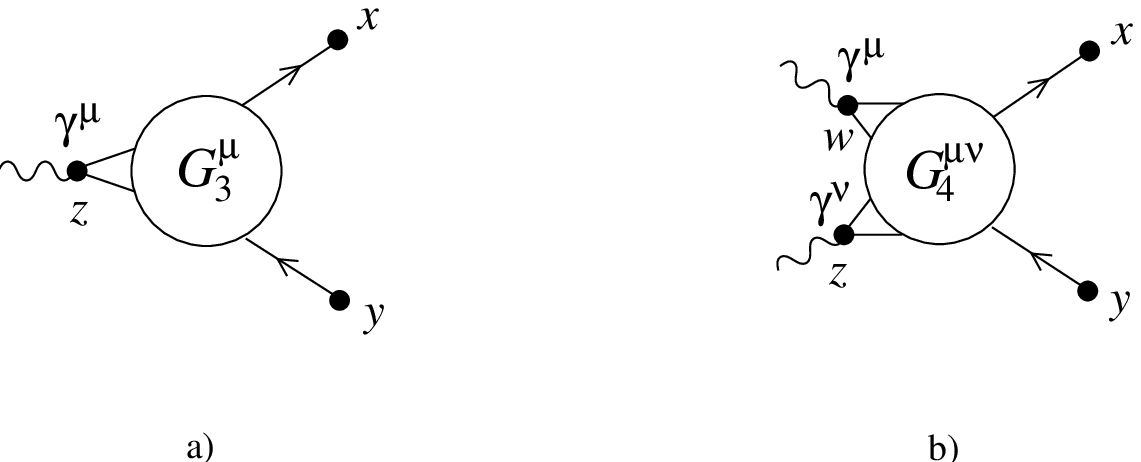}{Graphical representations of 
a) $G_3^{\mu}$ and b) $G_4^{\mu\nu}$ defined in \eqGF, 
where wavy line represents the external gauge field.}

Therefore the functional differentiation w.r.t. $A_\mu$ 
(and then setting $A=0$) of the SD eq.\eqSDeqlowest\ 
successively generates the Bethe-Salpeter 
(BS) equations for the $G_{n+2}^{\mu_1\cdots\mu_n}$ functions 
and they are automatically (external) gauge covariant. 
It is convenient to define the following vertex function by amputating 
the fermion legs:
$$
  \Gamma_{n+2}^{\mu_1\cdots\mu_n} \equiv 
  \Sfinv G_{n+2}^{\mu_1\cdots\mu_n} \Sfinv .
\eqno\eq
$$
To show what is going on as explicitly as possible, 
from here on in this section,  we confine ourselves to the simplest 
case using the lowest order kernel
($\kterm=\kterm^{(1)}$). 
The extension to more complicated case will be trivial. 
First differentiation $\delta /\delta A_\mu $ of \eqSDeqlowest\ gives 
(see Fig.~4) 
$$
  \Gamma_3^\mu = \gamma^\mu + \Ktil \ast \Gamma_3^\mu \ ,
\eqn\eqBSone
$$
where $\Ktil \ast \Gamma_3^\mu \equiv K \ast ( \Sf \Gamma_3^\mu \Sf)
= K \ast G_3^\mu $ 
is defined in the same way as in Eq.\eqDefKop.
\figinsert\figBSone{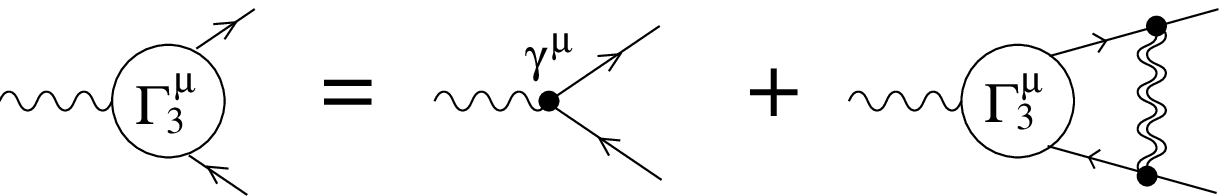}{BS equation for $\Gamma_3$.}
\noindent
Second differentiation of \eqSDeqlowest\ gives (see Fig.~5)
$$
   \Gamma_4^{\mu\nu}  
  -  \Gamma_3^\nu \Sf \Gamma_3^\mu  
  -  \Gamma_3^\mu \Sf \Gamma_3^\nu  
  =  \Ktil \ast \Gamma_4^{\mu\nu} \ .
\eqn\eqBStwo
$$
\figinsert\figBStwo{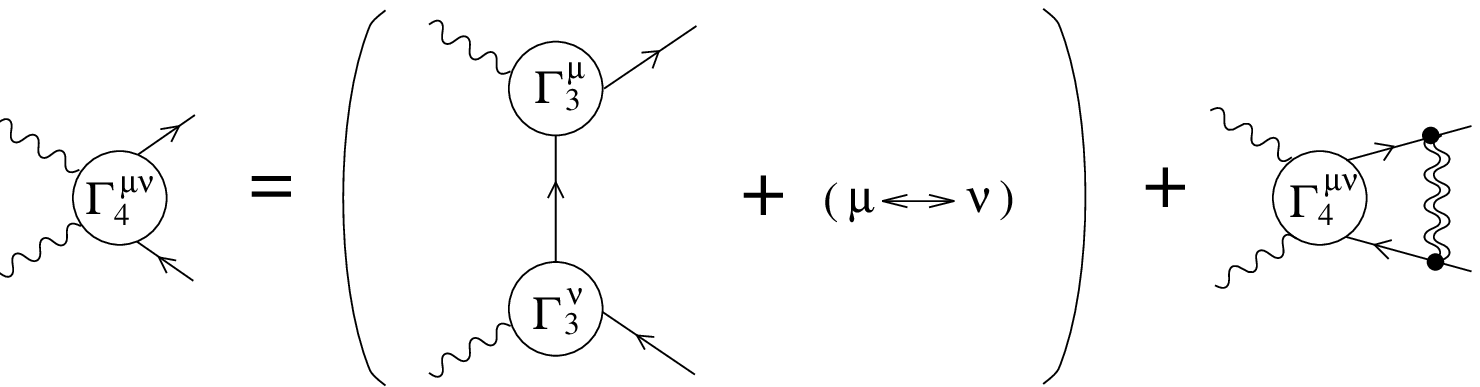}{BS equation for $\Gamma_4$.}
\noindent
Third differentiation of \eqSDeqlowest\ gives 
$$
\eqalign{
& 
  \Gamma_5^{\mu\nu\rho}  
  -
  \left(
    \Gamma_4^{\mu\nu} \Sf \Gamma_3^\rho  
    +
    \Gamma_3^\rho \Sf \Gamma_4^{\mu\nu} 
    + 
    \hbox{\rm ($\mu$, $\nu$, $\rho$ permutations)}
  \right)
\cr
& +
  \left(
    \Gamma_3^\mu \Sf \Gamma_3^\nu \Sf \Gamma_3^\rho 
    +
    \hbox{\rm ($\mu$, $\nu$, $\rho$ permutations)}
  \right) 
\cr
& \quad
  = \Ktil \ast \Gamma_5^{\mu\nu\rho} \ .
\cr
}
\eqn\eqBSthree
$$
We can see by the help of Eq.\eqBStwo\  that 
the last terms of the LHS in Eq.\eqBSthree\ 
play the role of cancelling the double counting of some diagrams 
contained in the second terms. 
Thus this equation reduces to
$$
\eqalign{
& 
  \Gamma_5^{\mu\nu\rho} 
  -
  \left(
     (\Ktil \ast \Gamma_4^{\mu\nu}) \Sf \Gamma_3^\rho 
    +
    \Gamma_3^\rho \Sf (\Ktil \ast \Gamma_4^{\mu\nu}) 
    + 
    \hbox{\rm ($\mu$, $\nu$, $\rho$ permutations)}
  \right)
\cr
& -
  \left(
    \Gamma_3^\mu \Sf \Gamma_3^\nu \Sf \Gamma_3^\rho 
    +
    \hbox{\rm ($\mu$, $\nu$, $\rho$ permutations)}
  \right) 
\cr
& \quad
  = \Ktil \ast \Gamma_5^{\mu\nu\rho} \ .
\cr
}
\eqn\eqBSthreeSub
$$

Eqs.\eqBSone, \eqBStwo\  and \eqBSthreeSub\ are the BS equations 
determining the vertices  
$\Gamma_3^{\mu}, \Gamma_4^{\mu\nu}$ and $\Gamma_5^{\mu\nu\rho}$, 
respectively, in the ladder approximation. 
It is in fact easy to find 
formal solutions to these equations. Define the four-point fermion 
Green function $L$ in the ladder approximation as given in Fig.~6.
\figinsert\figLDef{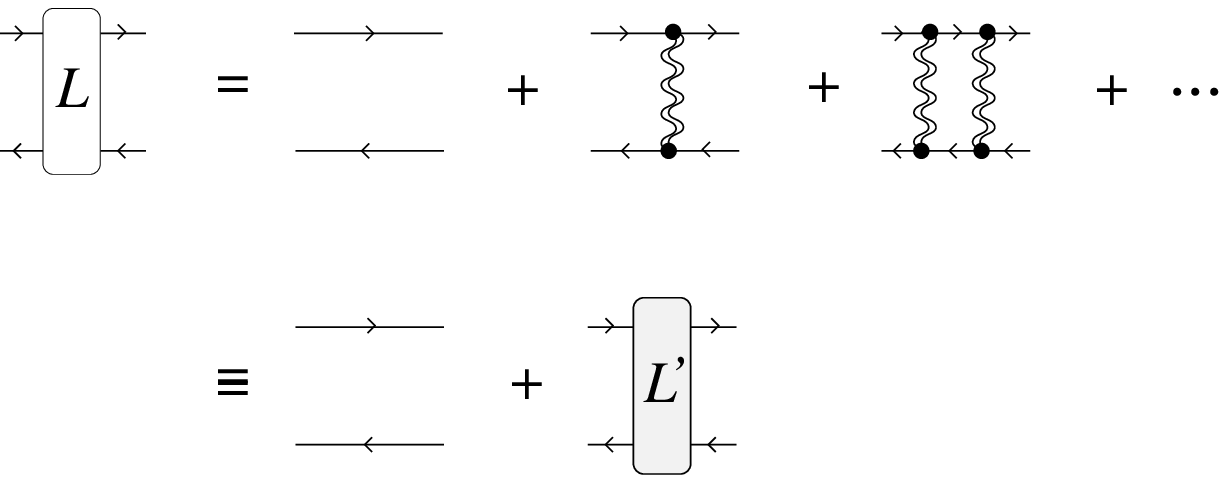}{Definition of $L$ and $L'$.}
\noindent
Then, inspection of the 
Eqs.\eqBSone, \eqBStwo\  and \eqBSthreeSub\ 
shows that the formal solutions for 
$\Gamma_3^{\mu}, \Gamma_4^{\mu\nu}$ and $\Gamma_5^{\mu\nu\rho}$ are
given by Figs.~7, 8 and 9, respectively. 
\ifnum\figcond=1 \ifnum\figsty=0 \vskip10pt \fi\fi
\figinsert\figFSolone{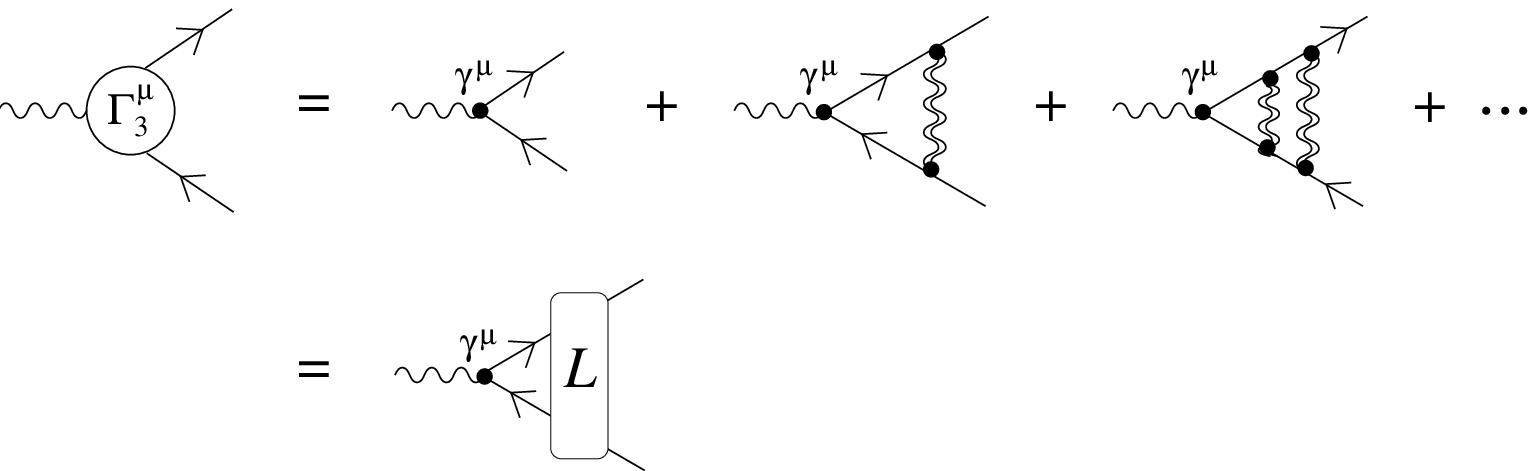}{Formal solution of $\Gamma_3^\mu$ 
using $\kterm=\kterm^{(1)}$.}
\ifnum\figcond=1 \ifnum\figsty=0 \vskip20pt \fi\fi
\figinsert\figFSoltwo{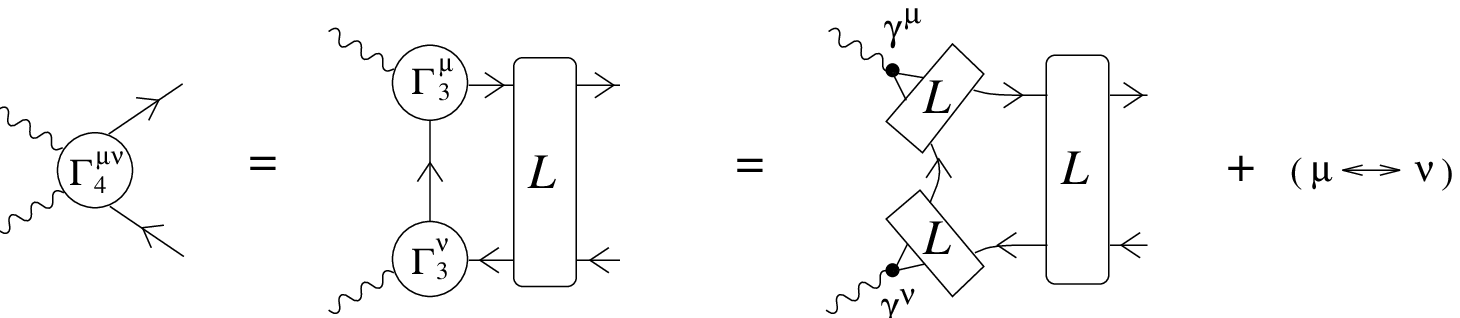}{Formal solution of 
$\Gamma_4^{\mu\nu}$ using $\kterm=\kterm^{(1)}$.}
\ifnum\figcond=1 \ifnum\figsty=0 \vskip20pt \fi\fi
\figinsert\figFSolthree{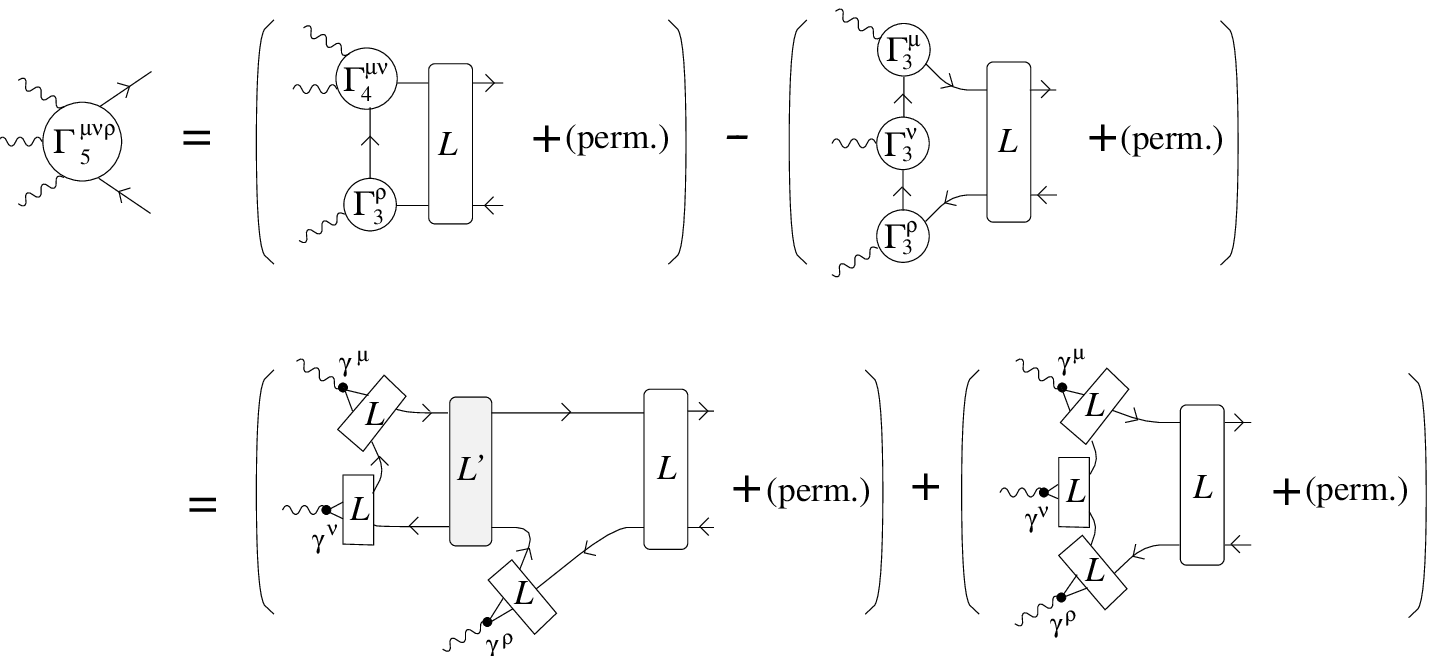}{Formal solution of 
$\Gamma_5^{\mu\nu\rho}$ using $\kterm=\kterm^{(1)}$.}

\ifnum\figcond>0 \ifnum\figsty=0 \ \vskip30pt \fi\fi

\chapter{Ward-Takahashi Identity}

We now show explicitly that the (external) gauge invariance of 
the effective action (or the covariance of our SD and BS equations) 
implies that the vertex functions determined by those BS equations 
satisfy Ward-Takahashi identities. 

Gauge invariance of the effective action $\Gamma[\Sf,A]$ implies
$$
  \Gamma[\Sf,A] = \Gamma[\Sf',A^U] 
\eqno\eq
$$
with $\Sf'\equiv U\Sf U^{-1}$,
which leads to the gauge covariance of the SD equation:
$$
  { \delta \Gamma[\Sf,A] \over \delta \Sf }
  =
  U^{-1}
  { \delta \Gamma[\Sf',A^U] \over \delta \Sf' }
  U \ .
\eqn\eqSDcov
$$
Therefore, if $\Sf=\Sf[A]$ is a solution to the SD equation 
on the background $A$, i.e.,
$$
  \left.
    { \delta \Gamma[\Sf,A] \over \delta \Sf }
  \right\vert_{\Sf=\Sf[A]}
  =0 \ ,
\eqno\eq
$$
then,
its gauge-transformed one $\Sf'=U\Sf[A]U^{-1}$ gives a solution
to the SD equation on the gauge-transformed background $A^U$:
$$
  \left.
    { \delta \Gamma[\Sf',A^U] \over \delta \Sf' }
  \right\vert_{\Sf'=U\Sf[A]U^{-1}}
  =0 \ .
\eqno\eq
$$
Namely, we have shown
$$
  \Sf[A^U] = U \Sf[A] U^{-1} \ .
\eqn\eqGTRSF
$$
If there are several solutions $\Sf^i[A]$ ($i=1,2,\cdots$) 
for a single background $A$,  we should have relation 
$\Sf^i[A^U] = U \Sf^i[A] U^{-1}$ for each $i$,
because of the continuity for $U\rightarrow1$.

Substituting the expansion \eqSfExpand\  into both sides of \eqGTRSF,
we have
$$
\eqalign{
  \hbox{LHS}
& =
  \Sf + i A_\mu^U G_3^\mu + 
  { i^2 \over 2 } A_\mu^U A_\nu^U G_4^{\mu\nu} + \cdots ,
\cr
  \hbox{RHS}
& =
  U \Sf U^{-1} + i A_\mu U G_3^\mu U^{-1} + 
  { i^2 \over 2 } A_\mu A_\nu U G_4^{\mu\nu} U^{-1} + \cdots \ .
\cr
}
\eqn\eqWTexpand
$$
In particular, for an infinitesimal gauge transformation 
$U=1+i \theta$ ($\theta =\theta^a \lambda^a$),
for which $A_\mu ^U$ is given by
$$
  A_\mu^{aU} = 
  A_\mu^a + D_\mu \theta^a=
  A_\mu^a + \delm \theta^a + f_{abc} A_\mu^b \theta^c 
\eqno\eq
$$
with structure constant $f_{abc}$ of the flavor group, 
we find
$$
\eqalign{
  \hbox{LHS}
& =
  \left(
    \Sf + i \delm \theta^a G_3^{a\mu} 
  \right)
  +  i A_{\mu}^a 
  \left[
    \left(
      \delta_{ab} + f_{acb} \theta^c
    \right)
    G_3^{b\mu} + i \deln \theta^b G_4^{a\mu,b\nu} 
  \right]
  + \cdots ,
\cr
  \hbox{RHS}
& =
  \Sf + i 
  \left(
    \theta \Sf - \Sf \theta
  \right)
  +  i A_{\mu}^a 
  \left[
    G_3^{a\mu} + i 
    \left(
      \theta G_3^{a\mu} - G_3^{a\mu} \theta
    \right)
  \right]
  + \cdots ,
\cr
}
\eqno\eq
$$
are equal with each other. 
Equating each power term in $A_\mu$ on both sides, we find
$$
\eqalignno{
  -i \delm^z G_3^{a\mu} (x,y;z) 
&=
  i \delta^4(z-x) \lambda^a \Sf(x-y) 
  - i \delta^4(z-y) \Sf(x-y) \lambda^a ,
&\eqname\eqWTone
\cr
  - i \deln^w G_4^{a\mu,b\nu} (x,y;z,w)
&=
  - f_{abc} \delta^4(w-z) G_3^{c\mu}(x,y;z)
& 
\cr
& ~~
  {}+ i \delta^4(w-x) \lambda^b G_3^{a\mu}(x,y;z)
  - i \delta^4(w-y) G_3^{a\mu}(x,y;z) \lambda^b ,
&\eqname\eqWTtwo
\cr
}
$$
$\ldots$, and so on. These are just the Ward-Takahashi identities 
required by the external gauge invariance. For instance, 
Eq.\eqWTone\ is nothing but the WT identity:
$$
  \delm^z 
  \bra{0}
    {\rm T} j^{a\mu} (z) \psi(x) \bar{\psi}(y)
  \ket{0}
  =
  - \delta^4(z-x) \lambda^a
  \bra{0}
    {\rm T} \psi(x) \bar{\psi}(y)
  \ket{0}
  + \delta^4(z-y)
  \bra{0}
    {\rm T} \psi(x) \bar{\psi}(y)
  \ket{0}
  \lambda^a \ .
$$
Thus this proves that the fermion 
propagator $\Sf$ and the vertices $\Gamma_{n+2}^{\mu_1\cdots\mu_n}$
determined by our SD and BS equations satisfy the Ward-Takahashi 
identities giving relations among them; namely, 
{\it our approximations for the SD and BS equations 
are mutually consistent and gauge invariant.}

Note that we have proven that this external gauge invariance holds 
not only for the simplest ladder case (i.e., with $\kterm^{(1)}$), 
but also for any order of approximations.\foot{
The gauge invariance for 
the 3-point vertex $\Gamma_3^{a\mu }$ in the simplest ladder case 
has been known for a long time to Maskawa and Nakajima\refmark{\MN}.
A refinement of the proof and the genelarization to the running
coupling case was given by Kugo and
Mitchard\refmark{\refKugoMitchard}.
}
For example, if we take $\kterm=\kterm^{(1)} + \kterm^{(2a)}$,
then the SD equation for $\Sf$ and the BS equation for 
$\Gamma_3^{a\mu }$ are changed into the forms as shown 
in Figs.~10 and 11.
These equations are much more complicated than the simple 
ladder ones, nevertheless they satisfy the gauge invariance. 
Important is the mutual consistency of the approximations 
between the SD equation and BS equations.
\figinsert\figHighSD{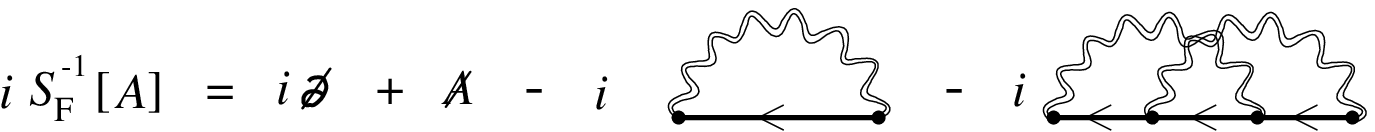}{SD equation using
$\kterm=\kterm^{(1)}+ \kterm^{(2a)}$.}
\figinsert\figHighBS{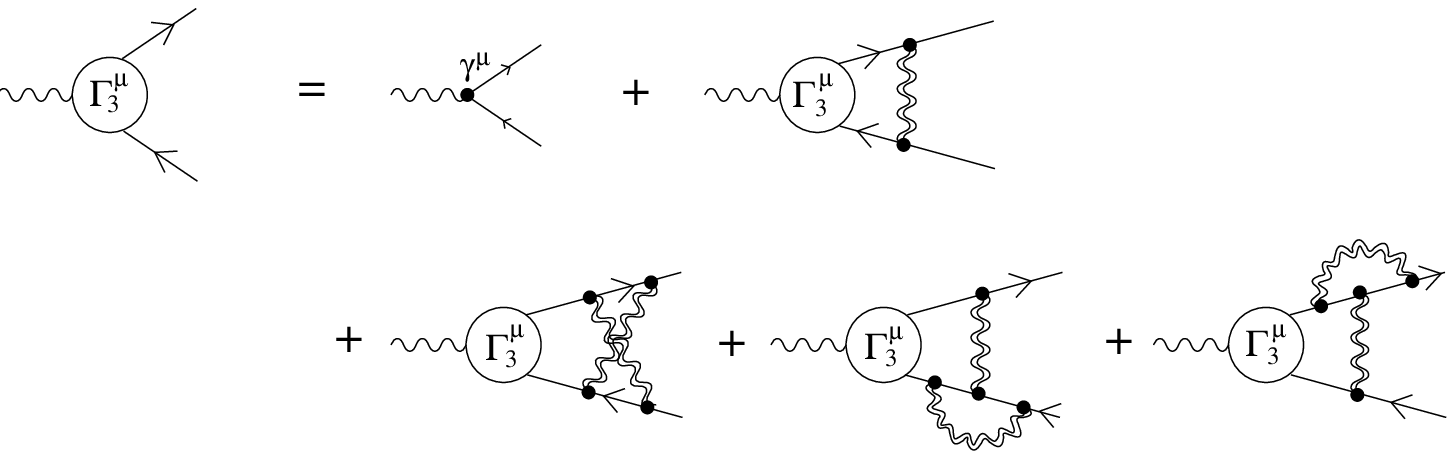}{BS equation for $\Gamma_3^\mu$
using $\kterm=\kterm^{(1)}+ \kterm^{(2a)}$.}

\ifnum\figcond>0 \ifnum\figsty=0 \vfill\eject \ \vskip1pt \fi\fi

\chapter{Anomaly}

We now turn to the problem of triangle anomaly. For definiteness, 
we work in $SU(3)_c$ QCD with two flavors; we consider 
$SU(2)_{\rm L}\times SU(2)_{\rm R}$ chiral 
symmetry limit in which the $u$ and $d$ quarks have zero masses. 

Consider fermion 2-point function with a vector and an axial-vector 
current inserted, $G_4^{Q\mu,\pi\alpha}$,
where $Q\mu$ denote the vector current to which photon couples,
$$
\eqalign{
  j^\mu 
& =
  \bar{q} Q \gamma^\mu q \ ,
\cr
& Q 
  = 
  \left(
    \matrix{
      {2\over3} & \cr
      & -{1\over3} \cr
    } 
  \right) 
  =
  \left(
    {1\over6} {\bf 1} + \pigen
  \right) ,
  \qquad
  q = 
  \left(
    \matrix{
      u \cr
      d \cr
    }
  \right) ,
\cr
}
$$
and $\pi\alpha$ denotes the axial-vector current to which pion 
$\pi ^0$ couples:
$$
  j_5^\alpha = 
  \bar{q} \pigen \gamma^\alpha \gamma_5 q \ .
$$

Then the WT identity Eq.\eqWTtwo, which was proven for the vector case 
in the previous section and can straightforwardly be extended to the 
axial-vector case, implies the following 
vector and axial-vector WT identity:
$$
\eqalignno{
&  -i \del_\mu^y G_4^{Q\mu,\pi\alpha}(w,z;y,x) \cr
& \qquad\qquad = 
  i \delta^4 (y-w)\, Q\, G_3^{\pi\alpha}(w,z;x) 
  - i \delta^4 (y-z)\, G_3^{\pi\alpha}(w,z;x)\, Q \ ,
& \eqname\eqVectorWT
\cr
&   -i \del_\alpha^x G_4^{Q\mu,\pi\alpha}(w,z;y,x) \cr
& \qquad\qquad = 
  i \delta^4 (x-w) \pigen\gamma_5 G_3^{Q\mu}(w,z;y) 
  + i \delta^4 (x-z) G_3^{Q\mu}(w,z;y)\, \pigen\gamma_5 \ ,
& \eqname\eqAxialWT
\cr
}
$$
where the $f_{abc}$ term does not appear because of $[Q,\tau_3/2]=0$.
Therefore, if we close the open fermion legs of 
$G_4^{Q\mu ,\pi \alpha }$ by inserting another vector vertex factor
$Q\gamma^\nu$,
we obtain {\it formally} both vector and axial-vector gauge invariant
3-point function:
$$
\eqalign{
  T^{\alpha\mu\nu}(x,y,z)
&\equiv
  i^3 e^2 \  \tr
  \left(
    G_4^{Q\mu,\pi\alpha}(z,z;y,x) Q \gamma^\nu
  \right) 
\cr
&=
  i^3 e^2 \  \tr
  \left(
    \bra{0}
      {\rm T} j^{\mu}(y) j_5^{\alpha}(x) \psi(z) \bar{\psi}(z)
    \ket{0}
    Q \gamma^\nu
  \right) 
\cr
&=
  - i^3 e^2 \,
  \bra{0}
    {\rm T} j_5^{\alpha}(x) j^{\mu}(y) j^{\nu}(z) 
  \ket{0} \ .
}
\eqn\eqTPT
$$
Indeed the vector and axial-vector gauge invariance of
$T^{\alpha\mu\nu}(x,y,z)$ is immediately 
seen from Eqs.\eqVectorWT\ and \eqAxialWT:
$$
\eqalignno{
&
  -i \delm^y T^{\alpha\mu\nu} (x,y,z)
& \cr
& \qquad = 
   e^2 \delta^4 (y-z) 
  \left[
    \tr
    \left(
      \left\{
        Q G_3^{5\alpha}(z,z;x) - G_3^{5\alpha}(z,z;x) Q  
      \right\}
      Q \gamma^\nu
    \right)
  \right]
& \cr 
& \qquad =0 \ ,
&\eqname\eqVecWT
\cr
&
  -i \del_\alpha^x T^{\alpha\mu\nu} (x,y,z)
& \cr
& \qquad =   
   e^2 \delta^4 (x-z) 
  \left[
    \tr
    \left(
      \left\{
        \pigen\gamma_5 G_3^{\mu}(z,z;y) + 
         G_3^{\mu}(z,z;y) \pigen\gamma_5 
      \right\}
      Q \gamma^\nu
    \right)
  \right] 
& \cr
& \qquad  =0 
  \quad \hbox{({\it formally})}\ .
&\eqname\eqAxWT
\cr
}
$$

We should note two things here:
\item{i)}
Despite its asymmetric looking in the above definition of 
$T^{\alpha \mu \nu }$, the two vector vertices $Q\mu$ and $Q\nu$ 
are in fact on the same footing.
Indeed, if the lowest order $\kterm^{(1)}$ is used, for instance, 
recall that the formal solution for $\Gamma_4$ was given in the form 
of Fig.~8.  Taking it into account also 
that $\Gamma _3$ was given by Fig.~7, 
we see that the amplitude $T^{\alpha \mu \nu }$ is written in the 
following 
manifestly symmetric form with respect to the two vector vertices 
and stands for ``triangle" diagram as shown in 
Fig.~12:
$$
  T^{\alpha\mu\nu}
  = i^3 e^2 \Tr
  \left(
    \Sf \Gamma^\mu \Sf \Gamma_5^{\alpha} \Sf \Gamma^\nu
  \right)
  + 
  \hbox{\rm ($\mu \leftrightarrow \nu$ cross term)} \ .
\eqn\eqTladder
$$
Here and henceforth, the vertices 
$\Gamma _3^{\pi \alpha }$ and $\Gamma _3^{Q\mu }$ are simply denoted as 
$\Gamma _5^{\alpha }$ and $\Gamma ^{\mu }$, respectively, for 
notational simplicity.
\figinsert\figTlet{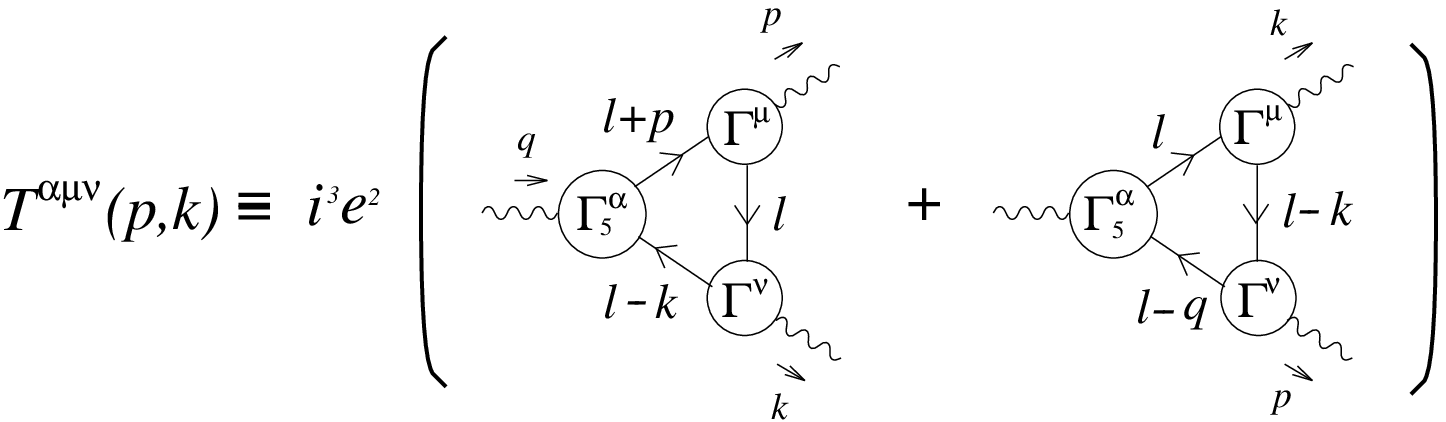}{Three point function in ladder
approximation.}
\item{}
Inspection shows that the same is also true for more general $\kterm$; 
for example, when we take $\kterm = \kterm^{(1)} + \kterm^{(2a)}$,
the formal solution for $T^{\alpha \mu \nu }$ 
take the form as given in Fig.~13, 
again showing manifest symmetry between the two vector vertices. 
\vskip 10pt
\figinsert\figTdeftwo{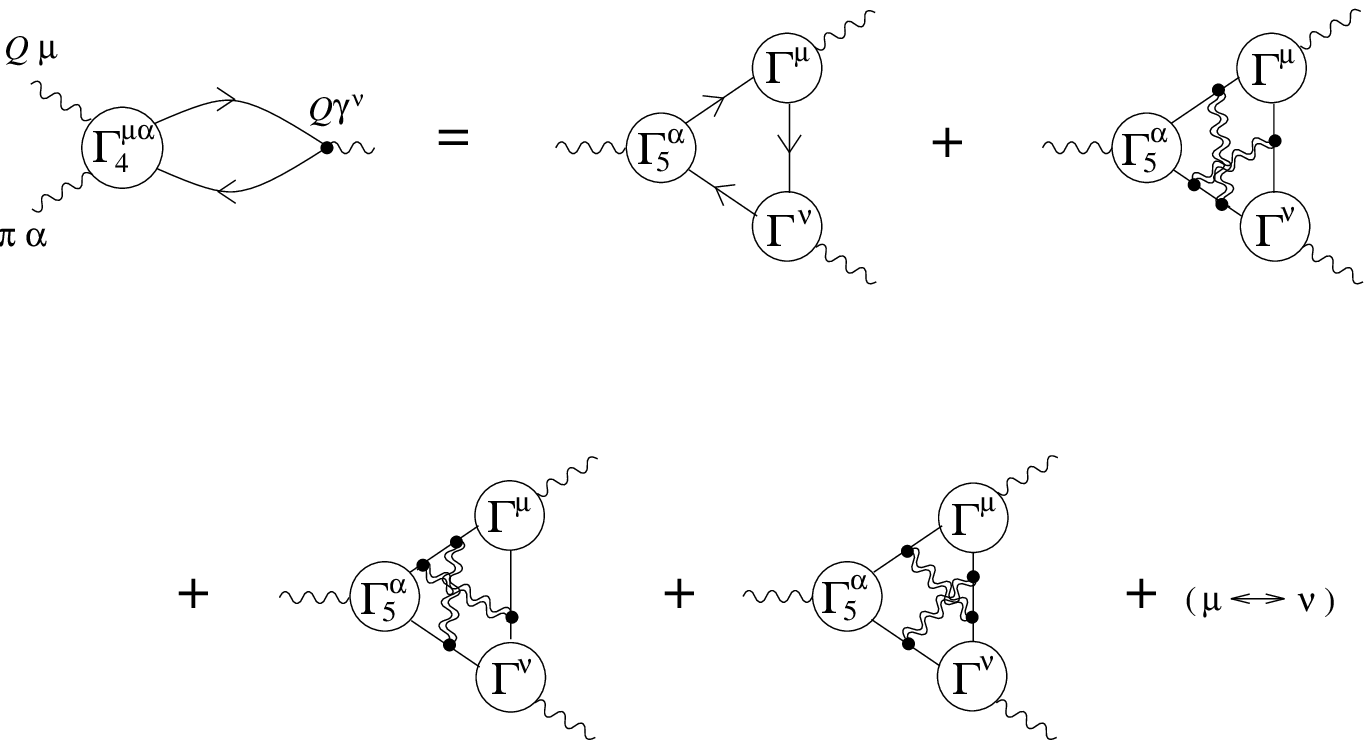}{Three point function 
using $\kterm = \kterm^{(1)} + \kterm^{(2a)}$.}

\item{ii)}
The above `proof' of the vector and axial-vector gauge invariance is 
just formal. 
The three point function $T^{\alpha \mu \nu }$ given in Eq.\eqTPT\ is 
{\it not} well-defined as it stands: the Green function 
$G_4(w,z;y,x)$ itself is well-defined but   
$G_4(z,z;y,x)$ at coincident point $w=z$, appearing after closing 
the fermion legs, is a divergent quantity and no longer well-defined.

The latter point is of course the well-known problem of anomaly. 
In order to define the three point function $T^{\alpha \mu \nu }$ 
properly, 
we need a regularization. Let us here adopt 
the Pauli-Villars-Gupta (PVG) regularization. (A brief 
discussion for the dimensional regularization is given in Appendix.) 
Since this regularization maintains the vector gauge-invariance,
we have only to examine the axial-vector WT identity:
$q_\alpha T^{\alpha\mu\nu} = ?$.
Hereafter we work with taking only the lowest order
$\kterm=\kterm^{(1)}$, for simplicity. 
Working in momentum space, we denote 
incoming momentum into the axial-vector vertex
$\Gamma_5^{\alpha}$ as $q_\alpha$, and out going momenta from the
vector vertices $\Gamma^\mu$ and $\Gamma^\nu$ as $p_\mu$ and
$k_\nu$, respectively. 
In PVG regularization, the regularized 3-point function 
$T^{\alpha\mu\nu}_{\rm(reg.)}$ is defined by
$$
\eqalign{
  T^{\alpha\mu\nu}_{\rm(reg.)}&(p,k) \cr
  \equiv & i^3 e^2
  \int \!\! {d^4\ell \over (2\pi)^4}
  \tr
  \biggl[
    \Sf(\ell) \Gamma^\mu \Sf(\ell+p)
    \Gamma_5^{\alpha} \Sf(\ell-k) \Gamma^\nu 
\cr
& \qquad\qquad\ \ \ \ \ \ \ 
    - \PVGSf(\ell)
    Q\gamma^\mu \PVGSf(\ell+p)
    \pigen\gamma^\alpha\gamma_5 \PVGSf(\ell-k)
    Q\gamma^\nu 
  \biggr]
\cr
& \qquad\qquad\ \ \ \ \ \ \ 
        + \hbox{\rm (cross term)}\  ,
\cr
}
\eqn\eqREGT
$$
where we have omitted the obvious momentum arguments of the vertices 
and $\PVGSf$ is the propagator of the PVG regulator field 
with mass $M$:
$$
  \PVGSf(\ell)
  \equiv
  {i \over \slash{\ell} - M} \ .
\eqno\eq
$$
To see that this is really regularized, 
we actually need the high momentum behavior of the 
solutions $\Sf$, $\Gamma ^\mu$, $\Gamma^\nu$ and $\Gamma_{5}^\alpha$.
Since the gluon propagator is well regularized as stated in 
Eq.\eqREGGLUON, the loop integrals in the SD and BS equations are 
sufficiently convergent. From those equations one can convince oneself 
by a simple power counting argument that the solutions approach 
the free ones as $\ell\ \rightarrow \ \infty $:
$$
\eqalign{
\Sf(\ell) \ &\longrightarrow \ \slash{\ell} + {\cal O}(\ell^{-1}) 
\ ,\cr 
\Gamma ^\mu  \ &\longrightarrow \ Q\gamma ^\mu  + {\cal O}(\ell^{-2})
\ ,\cr
\Gamma _5^\alpha  \ &\longrightarrow \ \pigen\gamma ^\alpha \gamma _5 + 
{\cal O}(\ell^{-2}) \ .
}\eqn\eqASYMPTOTIC
$$
Therefore Eq.\eqREGT\ gives a well regularized quantity: 
the degree of {\it superficial} divergence of the 
loop integral in this expression \eqREGT\ 
is now only logarithmic (actually, the integral is convergent) 
and therefore the integral is 
independent of the shift of the loop momentum $\ell$\refmark{\Jackiw},
contrary to the unregularized case. 
For definiteness, however, we choose 
the momentum assignment for the $\mu  \leftrightarrow\nu $ 
cross term as shown in Fig.~12. Then, we can rewrite Eq.\eqREGT\ 
into the form 
$$
\eqalign{
  T^{\alpha\mu\nu}_{\rm(reg.)}(p,k) =
  i^3 e^2 \Tr 
  \biggl[ \Big( & G_4^{Q\mu ,\pi \alpha }(\ell, \ell-k ; p, q) \cr
   &- \PVGSf(\ell) Q\gamma ^\mu \PVGSf(\ell+p)
       \pigen\gamma^\alpha\gamma_5 \PVGSf(\ell-k) \cr
   &- \PVGSf(\ell) \pigen\gamma^\alpha\gamma_5 
       \PVGSf(\ell-q) Q\gamma ^\mu \PVGSf(\ell-k) \Big) Q\gamma ^\nu 
\biggr]
\cr}\eqno\eq
$$
with functional trace $\Tr$ implying also the momentum integration 
$\int d^4\ell/(2\pi )^4$. 

Using the momentum space version of Eq.\eqAxialWT\ 
$$
\eqalign{
q_\alpha \,G_4^{Q\mu ,\pi \alpha }(\ell, \ell-k; p, q) &= 
 i \pigen\gamma_5\,G_3^{Q\mu }(\ell-q,\ell-k) 
    + G_3^{Q\mu }(\ell,\ell+p) i \pigen\gamma_5
\cr
&=
 i \pigen\gamma_5\,\Sf(\ell-q) \Gamma^\mu  \Sf(\ell-k) 
     + \Sf(\ell) \Gamma^\mu  \Sf(\ell+p) i\pigen\gamma_5 \cr
}\ee
$$
and an algebraic identity
$$
  \slash{q} \gamma_5 = \PVGSfinv(\ell+p) i\gamma _5  + 
    i\gamma _5 \PVGSfinv(\ell-k) + 2 M \gamma_5 \ ,
\eqn\eqALGEBRAICID
$$
we find (see Fig.~14)
$$
\eqalign{
  q_\alpha T^{\alpha\mu\nu}_{\rm(reg.)}&(p,k) 
\cr
  =   i^3 e^2
      &\Tr \Biggl[
      \Bigl(
        i \pigen\gamma_5 \Sf(\ell-q) \Gamma^\mu \Sf(\ell-k)
       -  i \pigen\gamma_5 \PVGSf(\ell-q) 
        Q\gamma^\mu \PVGSf(\ell-k) 
     \Bigr) Q \gamma^\nu   
\cr
     & \ \ +  \Bigl(
        \Sf(\ell) \Gamma^\mu \Sf(\ell+p) i\pigen\gamma_5 
         -  \PVGSf(\ell) Q\gamma^\mu \PVGSf(\ell+p) i \pigen\gamma_5 
      \Bigr) Q \gamma^\nu     \Biggr]
\cr
&
  - i^3 e^2 \lim_{M\rightarrow\infty} 
  2 M\, \Tr
  \Bigl(
    \pigen\gamma_5 \PVGSf(\ell-k)\gamma^\nu \PVGSf(\ell) 
    \gamma^\mu \PVGSf(\ell+p) +({\rm cross\ term}) 
  \Bigr) \ .
\cr
}
\eqno\eq
$$
\figinsert\figaxwt{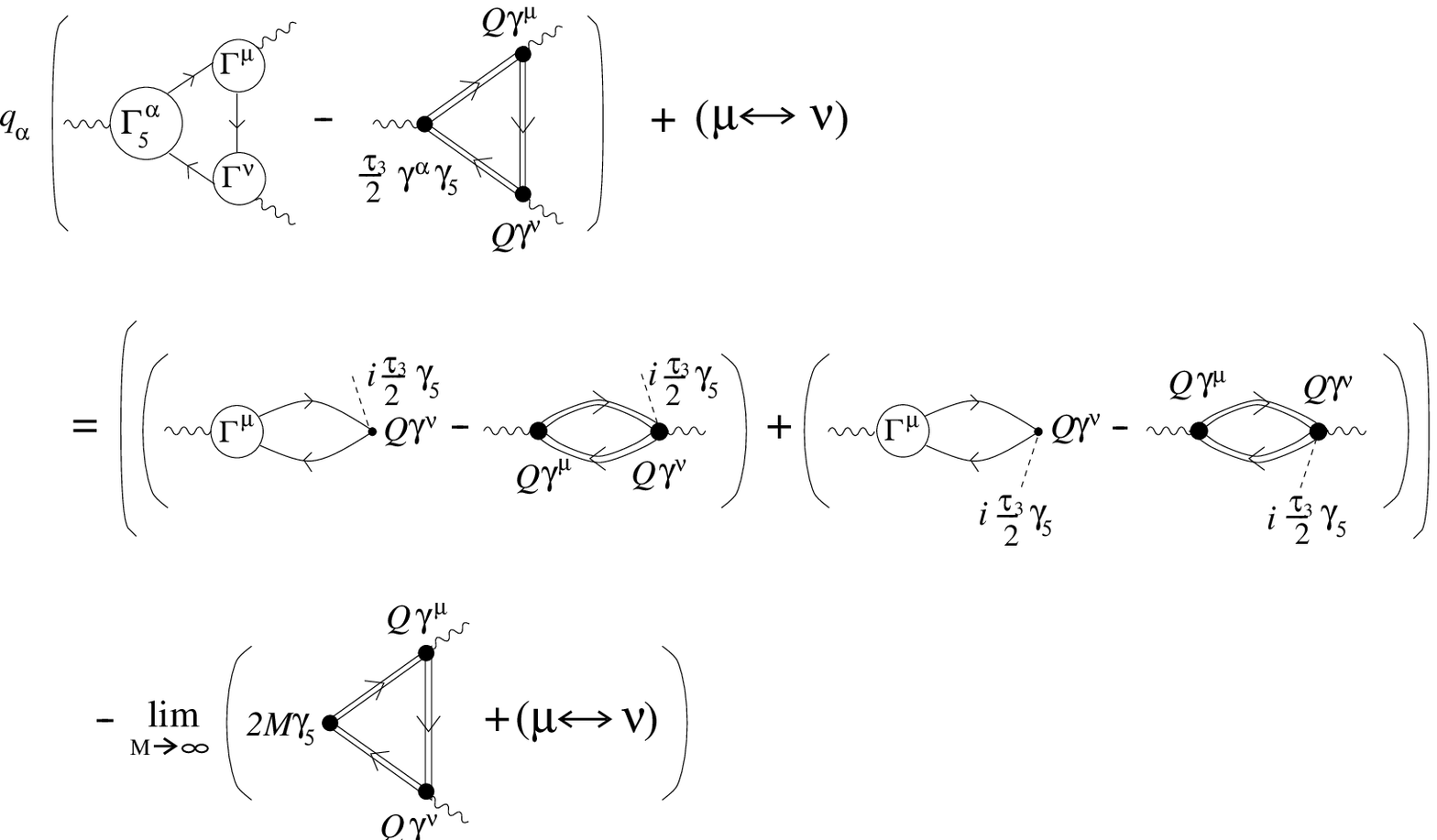}{%
Graphical representation of axial-vector Ward-Takahashi identity,
$q_\alpha T^{\alpha\mu\nu}_{\rm(reg.)}(p,k)$, where dashed lines
represent the insertion of $i\pigen\gamma_5$ and the place where the
momentum $q$ flows in, and the double solid lines represent the
propagator of PVG regulator field, $\PVGSf$.}
The first trace term in the RHS vanishes if 
we can shift the integration variable $\ell$ into $\ell+q$ 
for the two terms in the first line since, then, they take the same 
form as the two terms in the second line and cancel them exactly 
by $\{ i\pigen\gamma _5, Q\gamma ^\nu \}=0$. This shift of the 
integration variable 
separately for the first and the second term yields nonvanishing 
surface terms\refmark{\Jackiw}, 
which are, however, the same for both terms 
and cancel each other thanks to the asymptotic behavior \eqASYMPTOTIC\ 
of our solutions. 
The remaining last term reproduces the well-known anomaly:
$$
  q_\alpha T^{\alpha\mu\nu}_{\rm(reg.)}(p,k) 
    =   e^2 \AnomFac \equiv  \hbox{[Anomaly]} \ .
\eqn\eqANOMALY
$$

We thus have shown that our 3-point function calculated by using 
the nonperturbative fermion propagator $\Sf$ and vertices 
$\Gamma ^\alpha _5$ and $\Gamma ^\mu $, not only satisfy the 
conservation for the 
vector channels but does in fact reproduce correct anomaly for the 
axial-vector channel, the latter 
being in accord with Adler-Bardeen theorem\refmark{\refAdlerBardeen}. 
Note again that this nontrivial property is achieved only when 
the fermion propagator and the vertices are determined by such 
SD and BS equations with approximations 
mutually consistent with each other.

A comment may be in order:
although we have discussed only the 3-point function in this section,
it is clear that the same method as above can be applied to calculate
arbitrary $n$-point Green functions of current operators.
Namely, we can simply close the open fermion legs of the unamputated
vertex function $G_{(n-1)+2}^{\mu_1\cdots\mu_{n-1}}$ by inserting an
$n$-th vertex factor $\gamma^{\mu_n}$.
The reader can apply this procedure, for instance, to
$\Gamma_5^{\mu\nu\rho}$ given in Fig.~9 in the ladder approximation and
can see which types of the diagrams give gauge invariant 4-point Green
function for photons corresponding to
$\gamma\gamma\rightarrow\gamma\gamma$ scattering.
This example shows that even in the ladder approximation there
generally appear corrections intrinsic to the Green functions which
can be attributed neither to the propagator nor to the gauge boson
vertices.

\chapter{Low Energy Theorem and Pseudoscalar Bound State}

Thanks to these properties of our propagator and vertices, the decay 
amplitude for $\pi ^0 \rightarrow  2\gamma $, for instance, satisfies 
the low-energy theorem.  
This holds, of course, provided that we use 
BS amplitude for the boundstate $\pi ^0$ calculated by the `same' 
approximation as that used for propagator and vertices. To see this 
is the purpose of this section.

We consider the 3-point function 
$T^{\alpha\mu\nu}(p,k)$ for on-shell photon case, $p^2=k^2=0$, and 
recall the following fact which holds in the full theory and in 
our approximations at any order also. 
The vector gauge invariance 
($p_\mu T^{\alpha\mu\nu}(p,k)=k_\nu T^{\alpha\mu\nu}(p,k)=0$) 
and the bose symmetry $T^{\alpha\mu\nu}(p,k)=T^{\alpha\nu\mu}(k,p)$ 
together with odd parity property, determine 
the most general form of $T^{\alpha\mu\nu}$ as%
\foot{
Another possible form
$$
  A^{\alpha\mu\nu}(p,k)
  \equiv
  \epsilon^{\alpha\mu\nu\rho} (p_\rho - k_\rho)
  {q^2\over2} +
  \left(
    \epsilon^{\alpha\mu\rho\sigma} p^\nu 
    - \epsilon^{\alpha\nu\rho\sigma} k^\mu
  \right)
  p_\rho k_\sigma 
$$
can actually be expressed in terms of the above two amplitudes:
Indeed, from the identity 
$\epsilon^{[\mu\nu\rho\sigma}q^{\alpha]} = 0$,
we can prove
$$
  A^{\alpha\mu\nu}(p,k)
  =
  \epsilon^{\mu\nu\rho\sigma} p_\rho k_\sigma q^\alpha 
  -
  \left(
    \epsilon^{\alpha\mu\rho\sigma} k^\nu 
    - \epsilon^{\alpha\nu\rho\sigma} p^\mu
  \right)
  p_\rho k_\sigma  \ .
$$
}
$$
  T^{\alpha\mu\nu}(p,k)
  =
  \epsilon^{\mu\nu\rho\sigma} p_\rho k_\sigma q^\alpha F_1(q^2)
  +
  \left(
    \epsilon^{\alpha\mu\rho\sigma} k^\nu 
    - \epsilon^{\alpha\nu\rho\sigma} p^\mu
  \right)
  p_\rho k_\sigma  F_2(q^2) \ .
\eqn\eqGenForm
$$
Therefore generally we have 
$$
  q_\alpha T^{\alpha\mu\nu}(p,k)
  = 
  \epsilon^{\mu\nu\rho\sigma} p_\rho k_\sigma q^2 F_1(q^2)\ .
\eqn\eqMASSLESSPOLE
$$
The explicit $q^2$ factor in the RHS implies that 
{\it only the} ${1\over q^2}$ {\it pole term}
in $F_1(q^2)$ {\it can contribute} to this amplitude 
$q_\alpha T^{\alpha\mu\nu}$ in the limit of on-shell pion,
$q^2\rightarrow0$. (This is the reason why there exits a low-energy 
theorem to the $\pi ^0\rightarrow 2\gamma $ decay amplitude.) 
It is important to note that the vector gauge 
invariance is essential here to this conclusion; indeed, 
otherwise, the other invariant amplitude term of the form 
$ \epsilon ^{\alpha \mu \nu \rho }(p_\rho -k_\rho )F_3(q^2)$, for 
instance, could appear in 
Eq.\eqGenForm\ and contribute to this amplitude at $q^2=0$ 
{\it without} having massless pole.

We now also recall the definition of the BS amplitude: 
the BS amplitude $\chi $ as well as its conjugate $\bar\chi $ 
for the pion boundstate $\ket{\pi ({\bf q})}$ 
normalized invariantly as 
$\VEV{\pi ({\bf q})\vert\pi ({\bf q'})}
=(2\pi )^32\abs{{\bf q}}\delta ^3({\bf q}-{\bf q'})$, 
are defined by
$$
\eqalign{
\bra{0} {\rm T} \psi(x) \bar\psi(y) \ket{\pi ({\bf q})} 
&= e^{-iqX}\int {d^4p\over (2\pi )^4}\,e^{-ipr}\ 
\chi(p;q) \ ,
\cr
\bra{\pi ({\bf q})} {\rm T} \psi(y) \bar\psi(x) \ket{0} 
&= e^{iqX}\int {d^4p\over (2\pi )^4}\,e^{ipr}\ 
\bar\chi(p;q) \ , 
\cr}
\eqn\eqBSAMP
$$
where $X\equiv (x+y)/2$ and $r\equiv x-y$. 
It is easy to see that 
CPT invariance leads to the relation 
$\bar\chi (p;q) = \chi (p;-q)$,
because of which we often denote $\bar\chi$ as $\chi$ below for
notational simplicity.
This boundstate generally appears 
as a massless pole term in the 4-point amplitude 
$G_4\equiv \bra{0}{\rm T}\psi \bar\psi \psi \bar\psi \ket{0}$
in the form:
$$
G_4 = \chi\ {i\over q^2}\ \bar\chi  
+ \hbox{regular term at $q^2=0$}\ .
\eqn\eqGfour
$$
If we close the first $\psi $ and $\bar\psi $ legs of $G_4$ 
by inserting $\pigen\gamma^\alpha$, 
it becomes the axial-vector vertex $\Gamma _5^\alpha $ 
with 
fermion propagators attached. 
Therefore the pion pole appears in the 
axial-vector vertex in the form (see Fig.~15):
$$
  \Gamma_5^\alpha
  =
   i\, f_\pi q^\alpha {i\over q^2} \chihat 
  + \hbox{\rm (regular term)} \ ,
\eqn\eqPSpole
$$
where $\chihat$ is the amputated BS amplitude, 
$\chihat=\Sfinv\chi \Sfinv$, 
and the decay constant $f_\pi$ is given by
$$
\eqalign{
if_\pi q^\alpha 
&=\bra{0}\bar\psi (0){\tau _3\over 2}\gamma ^\alpha \gamma _5\psi 
(0)\ket{\pi ({\bf q})} \cr
&=
-\int {d^4p\over (2\pi )^4}\,\tr[ {\tau _3\over 2}\gamma ^\alpha \gamma 
_5
\chi(p;q) ] \ . \cr
}\eqn\eqFPI
$$
\figinsert\figaxbs{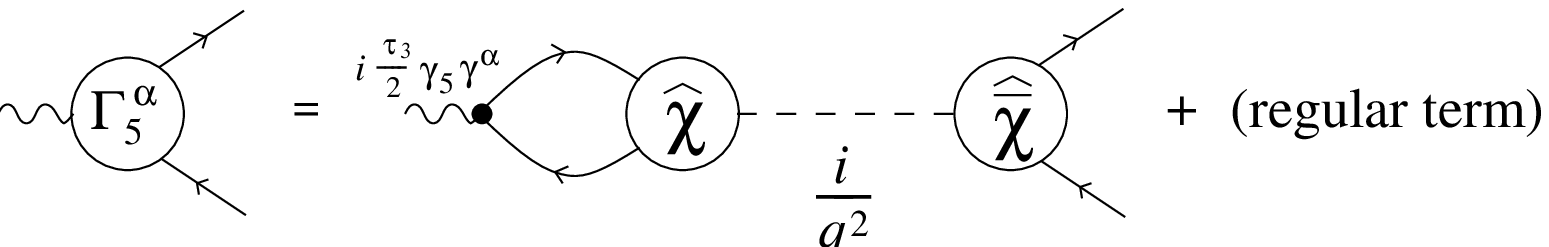}{BS amplitude of the pion, $\chihat$, 
in the axial-vector vertex, $\Gamma_5^\alpha$.}
\noindent
These are general arguments. If we use an approximation to determine 
$G_4$ and $\Gamma _5^\alpha $, the BS amplitude $\chi $ is also 
determined 
accordingly, via Eq.\eqGfour\ or \eqPSpole: for instance, if we use 
the ladder approximation ($\kterm=\kterm^{(1)}$), 
then the 4-point function $G_4$ is the ladder $L$ given in Fig.~6 
and therefore the BS amplitude $\chi $ should be determined by the 
homogeneous ladder BS equation as shown in Fig.~16. 
The decay constant $f_\pi $ is also the one calculated by Eq.\eqFPI\ 
using this BS amplitude. 
\figinsert\figBSpi{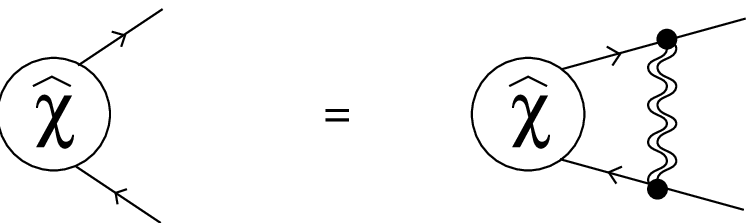}{%
Graphical representation of 
homogeneous ladder BS equation for $\chihat$.}

The massless pole in $T^{\alpha \mu \nu }$ required in 
Eq.\eqMASSLESSPOLE\ 
is provided by the pion pole in the axial-vector vertex $\Gamma 
_5^\alpha $ 
as given in \eqPSpole. In the ladder approximation, 
the 3-point function $T^{\alpha \mu \nu }$ is given in the form 
\eqTladder\ 
so that we find 
$$
  \lim_{q^2\rightarrow0} q_\alpha T^{\alpha\mu\nu}(p,k)
 =
  i e^2 f_\pi 
  \Tr
  \left(
    \chihat \Sf \Gamma^\nu \Sf \Gamma^\mu \Sf 
  \right)
  + \hbox{\rm (cross term)} \ ,
\eqno\eq
$$
which is shown in Fig.~17.
\figinsert\figAmplitude{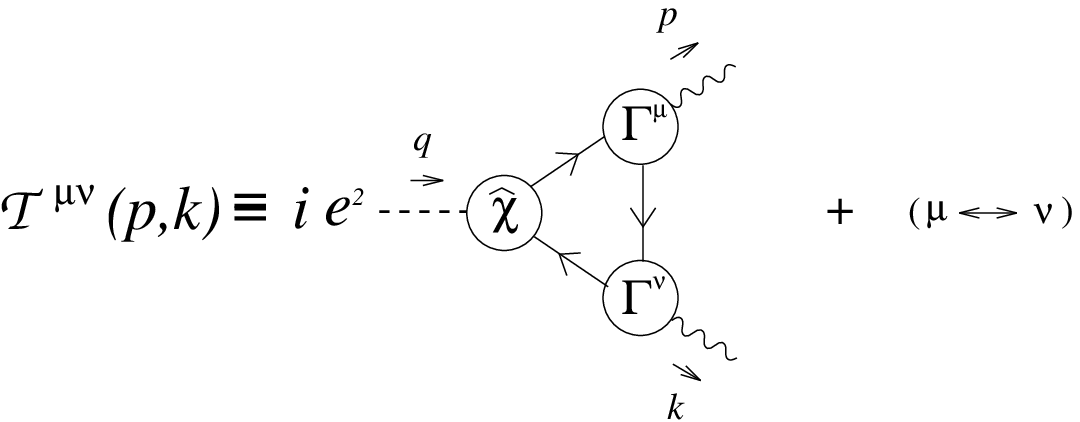}{%
Graphical representation of the function 
${\cal T}^{\mu\nu}(p,k)$.}
Combining with Eq.\eqANOMALY, this proves that 
$$
  {\cal T}^{\mu\nu}(p,k)
  \equiv
  i e^2
  \Tr
  \left(
    \chihat \Sf \Gamma^\nu \Sf \Gamma^\mu \Sf 
  \right)
  + \hbox{\rm (cross term)} 
  =
  {1\over f_\pi} \hbox{\rm [Anomaly]} \ .
\eqn\eqLET
$$
Namely, when the BS amplitude $\chi$ is calculated in the ladder
approximation,
then the $\pi^0\rightarrow2\gamma$ amplitude consistent with the low
energy theorem can be obtained by the LHS of Eq.\eqLET\ with vector
vertices and fermion propagator which are also calculated 
by the same order of approximation. It will also be easy to convince 
oneself that in the next order of approximation taking 
$\kterm = \kterm^{(1)} + \kterm^{(2a)}$, the correct amplitude 
satisfying the low-energy theorem can be obtained by calculating the 
diagrams given in the RHS of Fig.~13 with $\Gamma _5^\alpha $ vertex 
replaced 
by the BS amplitude $\chihat $.

\chapter{Saturation by zeroth order term}

In the ladder approximation,
there is a gauge in which the solution to the SD equation take the
form 
$$
  i \Sfinv(p) = \slash{p} - \Sigma(p^2) 
\eqn\eqPropL
$$
without wave function renormalization. 
[This is the case in Landau gauge for fixed coupling case%
\foot{If we use the regularized gluon propagator,
\eqREGGLUON,
there is a slight deviation from Landau gauge in the region
$p^2\gsim\Lambda^2$\refmark{\refKugoMitchard}.
},
and there is a non-local gauge realizing Eq.\eqPropL\ even for running
coupling case\refmark{\refKugoMitchard,\refGSC}.]
In such a case, there appears an interesting phenomenon that the
saturation of the low-energy theorem by the LHS of Eq.\eqLET\ 
is already realized by the {\it zeroth order terms in the external
momenta of the BS amplitude $\chihat$ and the vector vertices
$\Gamma^\mu$ and $\Gamma^\nu$.}
This is actually surprising because the linear terms in the external 
momenta individually give non-vanishing contributions to the amplitude.
The net effect of those linear terms vanishes.
Let us explain this fact.

First let us see what is the zeroth order term in the BS amplitude
$\chi$.
The WT identity Eq.\eqWTone\ for the axial-vector vertex reads
$$
  q_\mu \Gamma_5^\mu
  \left(
    p+{q\over2} , p-{q\over2}
  \right)
  =
  i \Sfinv \left(p+{q\over2}\right) \pigen \gamma_5
  + i \gamma_5 \pigen \Sfinv \left(p-{q\over2}\right) \ .
\eqno\eq
$$
Consider $q^\mu\rightarrow0$ limit on both sides:
For the LHS, only the $1/q^2$ pseudoscalar pole term given in 
Eq.\eqPSpole\ survives in this limit and yields
$$
  \lim_{q\rightarrow0}\hbox{\rm LHS}
  =
  - f_\pi \chihat (p;q=0)  \ .
\eqno\eq
$$
When $\Sf$ takes the form Eq.\eqPropL,
the RHS, on the other hand, gives
$$
  \lim_{q\rightarrow0}\hbox{\rm RHS}
  =
  - 2 \Sigma (p^2) \pigen \gamma_5 \ .
\eqno\eq
$$
So we find 
$$
  \chihat (p;q=0) =
  {2 \over f_\pi} \Sigma (p^2) \pigen \gamma_5 
  \equiv
  \chihat_0 (p) \ .
\eqno\eq
$$
This is an exact result giving the zeroth order term in $q$ as far as
$\Sf$ takes the form Eq.\eqPropL.
[Pagels-Stokar's proposal\refmark{\refPagelsStokar} 
is essentially to regard this as an
approximate BS amplitude up to order $q^2$:
$$
  \chihat_{\rm PS} (p;q) = 
  {2 \over f_\pi} \Sigma (p^2) \pigen \gamma_5 
  + {\cal O} (q^2) \ .
\eqno\eq
$$
As a matter of fact,
there appear three ${\cal O}(q)$ terms which generally take the form
$$
\eqalign{
  \chihat (p,q)
= &
  {2 \over f_\pi} \Sigma (p^2) \pigen \gamma_5 
  + {\chihat}_{\rm linear} + {\cal O}(q^2) \ ,
\cr
& 
  {\chihat}_{\rm linear} \equiv
  \left(
    \widehat{P}(p^2) ( p \cdot q ) \slash{p}
    + \widehat{Q} (p^2) \slash{q} 
    + \widehat{T}(p^2) {1\over2} 
    ( \slash{p} \slash{q} - \slash{q} \slash{p})
  \right)
  \gamma_5 \ .
\cr
}
\eqno\eq
$$
So the Pagels-Stokar approximation corresponds to discarding 
all those ${\cal O}(q)$ amplitudes.]

Next let us see the zeroth order form in the vector vertex $\Gamma^\mu$.
Again from the vector WT identity \eqWTone\ and the form \eqPropL\ 
of $\Sf$, we find
$$
\eqalign{
  p_\mu \Gamma^\mu 
  \left(
    \ell + {p \over 2} , \ell - {p \over 2}
  \right)
&=
  i \Sfinv \left( \ell + {p \over 2} \right) Q
  - Q i \Sfinv  \left( \ell - {p \over 2} \right)
\cr
&=
  p_\mu
  \left[
    \gamma^\mu - 2 \ell^\mu f (\ell^2,p\cdot\ell)
  \right] Q \ ,
\cr
}
\eqn\eqVecWTzero
$$
with 
$$
  f (\ell^2,p\cdot\ell)
  \equiv
  \sum_{k=0}^\infty
  {
    \left( p\cdot\ell \right)^2
  \over
    (2k+1) !
  }
  \left(
    { d \over d \ell }
  \right)^{2k+1}
  \Sigma(\ell^2) 
\eqno\eq
$$
for $p^2=0$ (on-shell photon).
This tells us that
$$
  \Gamma^\mu
  \left(
    \ell + {p \over 2} , \ell - {p \over 2}
  \right)
  =
  \left[ \gamma^\mu - 2 \ell^\mu f (\ell^2,p\cdot\ell)
  + \Gamma^\mu_{\rm tr.} \right] Q \ ,
\eqno\eq
$$
where $\Gamma^\mu_{\rm tr.}$ stands for transversal part satisfying 
$p_\mu \Gamma^\mu_{\rm tr.}=0$ by itself.
Since $\Gamma^\mu_{\rm tr.}$ generally takes the form
$$
\eqalign{
  \Gamma^\mu_{\rm tr.}
= &
  {1\over2} [ \gamma^\mu \slash{p} - \slash{p} \gamma^\mu ]
  A (\ell^2,p\cdot\ell)
  +
  \epsilon^{\mu\nu\rho\sigma} \ell_\nu p_\rho \gamma_\sigma \gamma_5
  B (\ell^2,p\cdot\ell)
\cr
& +
  ( p^2 \ell^\mu - p^\mu p\cdot\ell )
  \left[
    C_1 (\ell^2,p\cdot\ell) + \slash{p} C_2 (\ell^2,p\cdot\ell) 
    + \slash{\ell} C_3 (\ell^2,p\cdot\ell) 
  \right]
\cr
& +
  ( p^2 \gamma^\mu - \gamma^\mu p^2 )
  D (\ell^2,p\cdot\ell) + \cdots \ ,
\cr
}
\eqno\eq
$$
and is at least linear in the external momentum $p$,
the zeroth order term in $p$ of $\Gamma^\mu$ 
is given by
$$
  \left.
    \Gamma^\mu
    \left(
      \ell + {p \over 2} , \ell - {p \over 2}
    \right)
  \right\vert_{p=0}
  =
  \gamma^\mu - 2 \ell^\mu \Sigma'(\ell^2)
  \equiv
  \Gamma^\mu_0 ( \ell ) \ .
\eqno\eq
$$
Note that $\ell^\mu$ here stands for the average momentum of those of
incoming and outgoing fermions.

Let us now show that the zeroth order terms $\chihat_0$ and 
$\Gamma_0^\mu$ of the BS amplitude and the vector vertices already 
saturate the low-energy theorem;
namely, the equality
$$
\eqalignno{
  {\cal T}^{\mu\nu}(p,k)
= &
  {\cal T}^{\mu\nu}_0(p,k)
  =
  {1\over f_\pi} \hbox{\rm [Anomaly]} ,
&\eq
\cr
& {\cal T}^{\mu\nu}_0(p,k)
  \equiv
  i e^2 \Tr
  \left(
    \chihat_0 \Sf \Gamma^\nu_0 \Sf \Gamma^\mu_0 \Sf 
  \right)
  + \hbox{\rm (cross term)} 
&\eqname\eqZeroAmpDef
\cr
}
$$
holds for the on-shell photons and pion ($p^2=k^2=q^2=0$).
A key observation for this equality is that
$$
\eqalign{
  q_\alpha \pigen \gamma^\alpha \gamma_5
& =
  \left(
    \slash{\ell} + \slash{p} - \Sigma(\ell+p) 
  \right)
  \pigen \gamma_5
  + 
  \gamma_5 \pigen
  \left(
    \slash{\ell} - \kslash  - \Sigma(\ell-k) 
  \right)
\cr
& \qquad \qquad
  +
  \left(
    \Sigma(\ell+p) + \Sigma(\ell-k) 
  \right)
  \pigen \gamma_5
\cr
& =
  i \Sfinv (\ell+p) \pigen \gamma_5 
  + \pigen \gamma_5  i \Sfinv (\ell-k) 
  +
  \chihat_0 
  \left(
    \ell + {p-k \over 2}
  \right)
  + {\cal O} (q^2) \ .
\cr
}
\eqn\eqAxIdent
$$
Eliminating $\chihat _0$ using this identity, we find
$$
\eqalign{
  f_\pi \,{\cal T}^{\mu\nu}_0(p,k)
= &
  - i e^2 
  \Biggl[
    \Tr
    \left(
      i \pigen \gamma_5 \Sf \Gamma^\mu_0 \Sf \Gamma^\nu_0
    \right)
    +
    \Tr
    \left(
      \Sf \Gamma^\mu_0 \Sf i \pigen \gamma_5 \Gamma^\nu_0 
    \right)
\cr
& \qquad
    +
    \Tr
    \left(
      \Sf i \pigen \gamma_5 \Gamma^\mu_0 \Sf \Gamma^\nu_0 
    \right)
    +
    \Tr
    \left(
      \Sf \Gamma^\mu_0 i \pigen \gamma_5 \Sf \Gamma^\nu_0 
    \right)
  \Biggr]
\cr
& + i e^2
  \left[
    q_\alpha \Tr
    \left(
      \pigen \gamma^\alpha \gamma_5 \Sf \Gamma^\nu_0
      \Sf \Gamma^\mu_0 \Sf
    \right)
    + \hbox{\rm (cross term)}
  \right]
\cr
}
\eqn\eqZeroAmp
$$
since the last ${\cal O}(q^2)$ term in Eq.\eqAxIdent\ cannot contribute 
to the on-shell amplitude after the loop integration.
Although we have suppressed the momentum arguments here, it is 
important to 
note that the factor $\pigen \gamma_5$ indicates also the place at 
which the momentum $q$ flows in, so that it is {\it not} 
anticommutative with the vector vertex $\Gamma _0^\mu $ (or $\Gamma 
_0^\nu $) which 
has nontrivial dependence of the leg momenta.  
The LHS is well-defined since the loop integration is convergent 
because of the presence of the BS amplitude $\chihat_0$ which damps 
sufficiently rapidly. But the terms in the RHS are not well-defined 
separately. In order to make each term well-defined, we add to the RHS 
the identity for the PVG propagator $\PVGSf$,
$$
0 = -ie^2
  \Bigl( \PVGSf q_\alpha \pigen \gamma ^\alpha \gamma _5 \PVGSf 
       - i\pigen \gamma _5 \PVGSf - \PVGSf i\pigen \gamma _5 
    - 2M \PVGSf \pigen\gamma_5 \PVGSf \Bigr) 
     Q\gamma ^\nu \PVGSf Q\gamma ^\mu  
$$
(following from Eq.\eqALGEBRAICID\ ) and its $\mu \leftrightarrow\nu $ 
exchanged one. Then Eq.\eqZeroAmp\  reads: 
$$
\eqalign{
  f_\pi \,{\cal T}^{\mu\nu}_0(p,k)
= &
  - i e^2 
  \Biggl[
    \Tr
    \left(
      i \pigen \gamma_5 \Sf \Gamma^\mu_0 \Sf \Gamma^\nu_0
    \right)_{\rm PVG}
    +
    \Tr
    \left(
      \Sf \Gamma^\mu_0 \Sf i \pigen \gamma_5 \Gamma^\nu_0 
    \right)_{\rm PVG}
\cr
& \qquad
    +
    \Tr
    \left(
      \Sf i \pigen \gamma_5 \Gamma^\mu_0 \Sf \Gamma^\nu_0 
    \right)_{\rm PVG}
    +
    \Tr
    \left(
      \Sf \Gamma^\mu_0 i \pigen \gamma_5 \Sf \Gamma^\nu_0 
    \right)_{\rm PVG}
  \Biggr]
\cr
& + i e^2
  \left[
    q_\alpha \Tr
    \left(
      \pigen \gamma^\alpha \gamma_5 \Sf \Gamma^\nu_0
      \Sf \Gamma^\mu_0 \Sf
    \right)_{\rm PVG}
    + \hbox{\rm (cross term)}
  \right]
\cr
& + ie^2
  \lim_{M\rightarrow\infty} 
  \left[
    2 M \, \Tr
    \left(
      \pigen \gamma_5 \PVGSf 
      Q \gamma_\nu \PVGSf 
      Q \gamma_\mu \PVGSf 
    \right)
    + \hbox{\rm (cross term)}
  \right] ,
\cr
}
\eqn\eqZeroAmpII
$$
where each trace term with suffix PVG denotes the regularized one 
obtained by subtracting the same form term with replacements 
$\Sf \rightarrow  \PVGSf $, $\Gamma _0^{\mu (\nu )} \rightarrow  
Q\gamma ^{\mu (\nu )}$ made. 
For instance, the first term reads
$$
    \Tr
    \left(
      i \pigen \gamma_5 \Sf \Gamma^\mu_0 \Sf \Gamma^\nu_0
    \right)_{\rm PVG}
 = 
    \Tr
    \left(
      i \pigen \gamma_5 \Sf \Gamma^\mu_0 \Sf \Gamma^\nu_0
      - i \pigen \gamma_5 \PVGSf Q\gamma^\mu \PVGSf Q\gamma^\nu
    \right)\ .
\ee
$$
In this regularized form we can make the shift of loop integration 
variables freely (\ie, producing no surface term) as explained before. 
Then the first four trace terms in Eq.\eqZeroAmpII\ cancel 
among themselves and vanish: 
Indeed, in order for such `biangle' diagrams with $\gamma_5$ inserted
to give non-vanishing contribution,
at least four $\gamma$ matrices have to be picked up,
so that only the (momentum-independent) elementary vertex parts 
$Q\gamma^\mu$ and $Q\gamma^\nu$ in $\Gamma^\mu_0$ 
and $\Gamma^\nu_0$ can contribute.
But then, the factors $i\pigen \gamma_5$ anti-commute with 
$\Gamma^\mu_0$ or $\Gamma^\nu_0$ and those diagrams cancel with 
each other.
This is the case since the shift of integration variables is allowed 
by the above `regularization'.
But, this `regularization' introduced an extra contribution given by 
the last term in Eq.\eqZeroAmpII,  
which is precisely the anomaly required by the low-energy theorem.
So what now remains to be proved is that the second term
in Eq.\eqZeroAmpII\ vanishes:
$$
\eqalign{
    q_\alpha \, T_0^{\alpha\mu\nu} (p,k) &= 0 \cr
  T_0^{\alpha\mu\nu} (p,k)
  &\equiv
  i^3 e^2
  \Tr
  \left(
      \pigen \gamma^\alpha \gamma_5 \Sf \Gamma^\nu_0
      \Sf \Gamma^\mu_0 \Sf
  \right)_{\rm PVG}
  + \hbox{\rm (cross term)} \ .
\cr}
\eqn\eqZeroAmpLast
$$
This can be proven if we can show that the amplitude
$T_0^{\alpha\mu\nu}$ 
satisfies (vector) gauge invariance,
$p_\mu T_0^{\alpha\mu\nu} = k_\nu T_0^{\alpha\mu\nu} = 0$.
Indeed, then, together with the Bose symmetry,
$T_0^{\alpha\mu\nu}$ has to have the general form
Eq.\eqGenForm\ and hence $q_\alpha T_0^{\alpha\mu\nu}$ can be
nonzero at $q^2=0$ only when a massless pole $1/q^2$ term 
exists in the axial-vector channel.
But the diagram of $T_0^{\alpha\mu\nu}$ is clearly regular at $q^2=0$
since the fermion here carries non-zero mass function $\Sigma$,
and hence $ q_\alpha T_0^{\alpha\mu\nu}=0$ on the mass shell $q^2=0$.

To show the gauge-invariance of $T_0^{\alpha\mu\nu}$,
we note that we can replace the zeroth order vertex function 
$\Gamma^\mu_0$ and $\Gamma^\nu_0$ in Eq.\eqZeroAmpLast\ 
by the full non-transverse parts
$\gamma^\mu - 2 \ell^\mu f(\ell^2,p\cdot\ell)$ and 
$\gamma^\nu - 2 \ell^\nu f(\ell^2,k\cdot\ell)$,
respectively, defined in Eq.\eqVecWTzero,
since the differences are of ${\cal O}(p^2)$ and ${\cal O}(k^2)$
and vanish on mass shell ($q^2=p^2=k^2=0$) after loop integration.
Then, with these replacements performed,
we can use the vector WT identity Eq.\eqVecWTzero\ to obtain
$$
\eqalign{
  p_\mu T_0^{\alpha\mu\nu} (p,k)
= & \,\,
  i^3 e^2
  \left[
    \Tr
    \left(
      \pigen \gamma^\alpha \gamma_5 \Sf \Gamma_0^\nu \Sf i Q
    \right)
    -
    \Tr
    \left(
      \pigen i Q \gamma^\alpha \gamma_5 \Sf \Gamma_0^\nu \Sf
    \right)
  \right]
\cr
& \, -
  i^3 e^2
  \left[
    \Tr
    \left(
      \pigen \gamma^\alpha \gamma_5 \Sf \Gamma_0^\nu i Q \Sf
    \right)
    -
    \Tr
    \left(
      \pigen \gamma^\alpha \gamma_5 \Sf i Q \Gamma_0^\nu \Sf
    \right)
  \right] .
\cr
}
\eqno\eq
$$
Here $Q$ indicates also the position from which the momentum $p$ 
flows out.  
The first two terms clearly cancel with each other.
The second two terms also cancel since, again, only the
(momentum-independent) elementary vertex part $Q\gamma^\nu$ in
$\Gamma^\nu_0$ can contribute because of the presence of $\gamma_5$
in the trace and hence the factor $Q$, indicating also the 
momentum-flow position, becomes commutative with $\Gamma^\nu_0$.
This finish the proof.

It is interesting to see explicitly how the diagram 
\eqZeroAmpDef\ with lowest order terms
$\chihat_0$, $\Gamma_0^\mu$ and $\Gamma_0^\nu$ inserted actually 
reproduce the value of the low-energy theorem:
$$
\eqalign{
  {\cal T}^{\mu\nu}_0(p,k)
= &
  -i e^2
  \int \!\! {d^4\ell \over i(2\pi)^4}
  \Biggl\{
    \tr
    \Biggl[
      \chihat_0 \left( \ell + {p-k\over2} \right)
      {1 \over \Sigma(\ell-k) - \slash{\ell} + \kslash  }
      \Gamma_0^\nu \left( \ell - {k\over2} \right)
\cr
& \qquad\qquad\qquad \qquad \times 
      {1 \over \Sigma(\ell) - \slash{\ell} }
      \Gamma_0^\mu \left( \ell + {p\over2} \right)
      {1 \over \Sigma(\ell+p) - \slash{\ell} - \slash{p} }
    \Biggr]
\cr
& \qquad\qquad +
    (p\leftrightarrow k,\,\mu\leftrightarrow\nu)
  \Biggr\}
\cr
= &
  e^2 \AnomFac { 2 \over f_\pi }
  \int d x 
  {
    x \Sigma(x) \left( \Sigma(x) - 2 x \Sigma'(x) \right)
  \over
    \left( \Sigma^2(x) + x \right)^3
  } \ ,
\cr
}
\eqno\eq
$$
where $x=-\ell^2$ denotes the Euclidean momentum.
Interestingly enough, the value of the integral in the last line 
is ${1\over2}$ {\it irrespectively} of the detailed functional form of
(nonzero) $\Sigma(x)$. This can easily be 
seen if one can make a change of the 
integration variable $x$ into dimensionless one $y=x/\Sigma^2(x)$. 
Thus the lowest order terms correctly reproduce 
the value of the low-energy theorem.

\vskip 30pt

\ACK
T.K. is supported in part by the Grant-in-Aid for Scientific Research
(\#04640292) from the Ministry of Education, Science and Culture.

\APPENDIX{A}{A.\ Dimensional Regularization}

If we use the dimensional regularization instead of PVG
regularization, the identity used in the algebraic manipulation
deriving the WT identity Eq.\eqAxWT\ is changed to
$$
\eqalign{
  \slash{q} \gamma_5
= &
  \left[
    \left(
      \slash{\ell} + {\slash{q}\over2}
    \right)
    -
    \left(
      \slash{\ell} - {\slash{q}\over2}
    \right)
  \right]
  \gamma_5
\cr
= &
  \left[
    \slash{\ell} + {\slash{q}\over2}
    - \Sigma
    \left(
      \left(
        \ell + {q\over2}
      \right)^2
    \right)
  \right]
  \gamma_5
    -
  \gamma_5
  \left[
    \slash{\ell} - {\slash{q}\over2}
    - \Sigma
    \left(
      \left(
        \ell - {q\over2}
      \right)^2
    \right)
  \right]
\cr
& \qquad
  + 
  \left[
    \Sigma
    \left(
      \left(
        \ell + {q\over2}
      \right)^2
    \right)
    - \Sigma
    \left(
      \left(
        \ell - {q\over2}
      \right)^2
    \right)
  \right]
  \gamma_5
  - 2 \slash{l} \gamma_5 \ ,
\cr
}
\eqno\eq
$$
where $\ell$ is the (now $n$-dimensional) loop momentum and $l$ is its
extra dimensional part ($\ell=\ell^{(4)}+l$).
The first 3 terms in the RHS is the usual one and the usual proof of
the WT identity applies to those terms:
then, they vanish since now the integral is regularized and the shift
of the integration variable is justified.
So we have to evaluate only the contribution from the last term coming
from the extra dimension part $l$:
$$
  q_\alpha T^{\alpha\mu\nu} (p,k)
  =
  2 i e^2 l_\alpha 
  \Tr 
  \left(
    \Gamma_5^{\alpha} \Sf \Gamma^\nu \Sf \Gamma^\mu \Sf
  \right)
  + \hbox{\rm(cross term)} \ .
\eqno\eq
$$

Since $l \propto n-4$, only the divergent parts can contribute in the
RHS. Since the propagator $\Sf$ and the vertices $\Gamma ^\mu$,
$\Gamma^\nu$ and $\Gamma_5^\alpha$
asymptotically behave as free one as explained in \eqASYMPTOTIC, 
we can evaluate as follows:
$$
  \hbox{\rm(RHS)}
= 
  \lim_{n\rightarrow4} 
  2 i e^2 
  \Tr
  \left(
    \slash{l} \gamma_5 {\Sf}_{\Sigma=0} 
    Q\gamma_\nu {\Sf}_{\Sigma=0}  Q\gamma_\mu {\Sf}_{\Sigma=0} 
  \right) \ .
\eqno\eq
$$
But this is again the same diagram which leads to the well-known
anomaly.

\refout

\vfill\eject

\ifnum\figsty>0 \parindent=25pt \figout \vfill\eject 
\ifnum\figcond>0 \figepsfout \fi \fi

\bye